\documentclass[preprint,12pt]{elsarticle}

\usepackage{graphicx}

    \usepackage[firstpage]{draftwatermark} 
    \SetWatermarkScale{6}
    \SetWatermarkLightness{0.9}

\usepackage{amssymb}

\usepackage{lineno}

\usepackage{hyperref}

\usepackage{xcolor}

\usepackage[utf8]{inputenc}
\usepackage{newunicodechar}

    \usepackage{longtable} 

\newunicodechar{−}{-}

\usepackage[a4paper, total={6.5in, 8.5in}]{geometry} 
\usepackage{multirow}
\usepackage{float}

\usepackage{subfigure}

\usepackage{comment}

\journal{Journal of Systems and Software}

\begin{document}

\begin{frontmatter}

\begin{titlepage}

  \newcommand{\HRule}{\rule{\linewidth}{0.5mm}} 

  {\center 


  \textsc{\Large This is a preprint of a paper accepted in:}\\[0.5cm] 
  \textsc{\large Journal of Systems and Software }\\[0.5cm] 


  \HRule \\[0.4cm]
  { \huge \bfseries Test Automation Maturity Improves Product Quality --\\ Quantitative Study of Open Source Projects Using Continuous Integration}\\[0.4cm] 
  \HRule \\[0.5cm]



  {\em\Large\textbf Authors:}\\
  \vspace{.2 cm}
  Yuqing Wang (Corresponding author)\\
  yuqing.wang@oulu.fi,\\
  M3S, Univeristy of Oulu \\
  
  \vspace{.5 cm}
  Mika V. M{\"a}ntyl{\"a}\\
  mika.mantyla@oulu.fi\\
  M3S,Univeristy of Oulu\\
  
  \vspace{.5 cm}
  Zihao Liu\\
  zihao.liu@student.oulu.fi \\
  M3S,Univeristy of Oulu\\
  
  \vspace{.5 cm}
  Jouni Markkula\\
  jouni.markkula@oulu.fi \\
  M3S,Univeristy of Oulu\\
  
    \vspace{.5 cm}
 }  







\vfill 

\end{titlepage}


\newpage

\title{Test Automation Maturity Improves Product Quality --\\ Quantitative Study of Open Source Projects Using Continuous Integration}



\author[inst1]{Yuqing Wang*}\cortext[cor1]{Corresponding author} \ead{yuqing.wang@oulu.fi}
\author[inst1]{Mika V. M{\"a}ntyl{\"a}} \ead{mika.mantyla@oulu.ﬁ}
\author[inst1]{Zihao Liu}
\ead{zihao.liu@student.oulu.ﬁ}
\author[inst1]{Jouni Markkula}
\ead{jouni.markkula@oulu.ﬁ}

\affiliation[inst1]{organization={M3S research unit, University of Oulu},
            addressline={Pentti Kaiteran katu 1}, 
            city={Oulu},
            postcode={90014}, 
            country={Finland}}

\begin{abstract}
 The popularity of continuous integration (CI) is increasing as a result of market pressure to release product features or updates frequently. The ability of CI to deliver quality at speed depends on reliable test automation. In this paper, we present an empirical study to observe the effect of test automation maturity (assessed by standard best practices in the literature) on product quality, test automation effort, and release cycle in the CI context  of open source projects. We run our test automation maturity survey and got responses from 37 open source java projects. We also mined software repositories of the same projects. 
 The main results of regression analysis reveal that, higher levels of test automation maturity are positively associated with higher product quality (p-value=0.000624) and shorter release cycle (p-value=0.01891); There is no statistically significant evidence of increased test automation effort due to higher levels of test automation maturity and product quality. Thus, we conclude that, a potential benefit of improving test automation maturity  (using  standard  best  practices) is product quality improvement and release cycle acceleration in the CI context of open source projects. We encourage future research to extend our findings by adding more datasets with different programming languages and CI tools, closed source projects, and large-scale industrial projects. Our recommendation to practitioners (in the similar CI context) is to utilize standard best practices to improve test automation maturity. 
\end{abstract}


\end{frontmatter}


\newcommand{\OurRQ}{Do higher levels of test automation maturity lead to better product quality without increased test automation effort and release time in the CI context of open source projects?}

\newcommand{\HoneContent}{Higher levels of test automation maturity are associated with higher product quality}

\newcommand{\HtwoContent}{Higher levels of test automation maturity are associated with increased test automation effort}

\newcommand{\HthreeContent}{Higher product quality is associated with increased test automation effort}

\newcommand{\HfourContent}{Higher levels of test automation maturity are associated with longer release cycles}

\newcommand{\HfiveContent}{Higher product quality is associated with longer release cycles}

\newcommand{\HsixContent}{Increased test automation effort is associated with longer release cycles}

\section{Introduction}
\label{Sec:intr}
\noindent 
Changing customer requirements and emerging technologies are driving the frequent changes to software products \cite{fitzgerald2017,karvonen2017}. Software organizations must rapidly release product features or updates on a continuous cycle \cite{fitzgerald2017}. Continuous Integration (CI) is widely adopted to enable rapid and frequent releases~\cite{staahl2014modeling}. CI is a software development practice that requires developers of the team to frequently integrate code changes \cite{fowler2006}.  According to the official website of Travis CI\footnote{https://travis-ci.org/} (a popular CI server tool), \textit{``Over 900k open source projects and 600k users are testing on Travis CI"}. Software projects and users on other popular CI server tools like Jenkins and CircleCI also accumulate to a considerable amount. Based on the statistic \cite{market19}, ``\textit{CI global market size was USD 402.8 million in 2017 and is expected to reach USD 1,139.3 million by 2023, at a compound annual growth rate of 18.7\%}''.

Test automation has a history of over three decades, since around 1990 \cite{garousi2017taNot}. After CI was introduced in 2006, test automation became the backbone of CI \cite{staahl2014modeling,fowler2006}. CI requires that each integration of code changes must be verified by a build that automatically executes tests to detect defects early \cite{fowler2006}. However, based on many sources (e.g., \cite{ghaleb2019empirical, rausch2017,staahl2017continuity,hilton2016usage}), in the CI context, immature test automation practices can lead to negative outcomes, e.g., ineffective in defecting integration effects, cost and schedule overruns, and slow feedback loop, thus, product quality suffers and releases delay - the failure of CI. Maturing test automation practices is essential for the success of CI in the current software industry \cite{shahin2017continuous,staahl2017continuity}. 
The software industry and research community refer to immature test automation practices as a lack of test automation maturity \cite{fewster1999software,icsoft20,worldqualityreport20}.
Prior scholars described that, at the high level of test automation maturity, ``\textit{test automation practices are defined, managed, measured, controlled, and effective within the organization}''~\cite{eldh2014towards,icsoft20,garousi2017taNot,fewster1999software}. The existing literature has presented a set of standard best practices to guide organizations to reach the high level of test automation maturity \cite{fewster1999software,wang2019,wang2018test}. For instance, in many test maturity models (such as TMap \cite{vroon2013tmap}, TestSPICE \cite{TestSPICE}, and TAIM \cite{eldh2020test}), an example best practice is -  define a test automation strategy to conduct test automation activities under given boundary conditions. 

However, an essential issue is whether standard test automation best practices (in the literature) would fit the CI context and enable CI success. Essentially, successful CI practices can deliver high quality products fast on a continuous cycle with reasonable costs. That is, to enable CI success, by using standard best practices in the literature, high product quality should be achieved without the expense of long release time and high costs. Yet, research on that is limited, as CI success was actively researched from CI process related factors (e.g., integration frequency, integration serialization and batching, building status communication \cite{staahl2014,ghaleb2019empirical,staahl2017continuity}) but few from a test automation maturity based view. As such, many CI practitioners are staying tuned and waiting for further evaluation before they actually use standard best practices to assess and improve their test automation practices~\cite{worldqualityreport20,wendler2012maturity,zampetti2020empirical,PractiTest2020}.

In this paper, we present an empirical study to observe the effect of test automation maturity (assessed by standard best practices in the literature) on product quality, test automation effort, and release cycle in the CI context of open source projects. The CI context there refers to the context in where test automation is adopted with CI tools. Our study observed test automation effort as the combination of test automation development effort and execution effort. The reason is, based on practitioner surveys \cite{hilton2016usage,infoworld}, approximately 60\% of CI costs are spent on test automation development and execution, while these two types of effort together influence the release time in CI practices. Our research question is:

\begin{itemize}
    \item \textbf{Research question:} \OurRQ 
\end{itemize}

To answer our research question, we built a conceptual model, using literature, to hypothesize the relationships among our observed variables: Test automation maturity, Product quality, Test automation effort, and Release cycle. From over 30k open source java projects, we selected 149 ones and sent out the Test automation maturity survey to their main contributors to explore the state of practice of test automation maturity in their CI context. We studied 37 projects, which have main contributors answered our survey, by analyzing survey responses and mining their repositories (GitHub repositories, CI repositories, issue tracker systems) to get metric data. Our dataset contains test automation  maturity survey responses, project data, test suite, and test logs of these 37 projects. That is made publicly available (https://doi.org/10.5281/zenodo.5831609). Using the quantitative evidence, we made the following observations: 
\begin{itemize}
\setlength\itemsep{0em}
    \item  Higher levels of test automation maturity are highly significantly associated with higher product quality. The effect of test automation maturity is more significant than product size, product complexity, product popularity, product age, team size, and integration frequency on product quality. Older products exhibit lower product quality.
    
    \item Increased test automation effort caused by improving  the level of test automation maturity and product quality is not evidenced. The effect of product complexity and team size is more significant than test automation maturity and product quality on test automation effort. More complex products spent less test automation effort. Larger teams spend more test automation effort.

    \item Higher levels of test automation maturity are significantly associated with shorter release cycles. Test automation effort does not seem to impact release time. The direct effect of product quality on release time is negative.
 
\end{itemize}

Our observations show that, in the CI context of observed open source projects, a potential benefit of improving the level of test automation maturity (using standard best practices in the literature) is product quality improvement and release cycle acceleration, while increased test automation effort caused by improving the level of test automation maturity and product quality is not evidenced. This suggests that using standard best practices in the literature to improve test automation maturity can enable CI success of open source projects. 

The remainder of this paper is structured as follows. Section~\ref{Sec:back} reviews related work. Section~\ref{Sec:meth} describes our research method. Section~\ref{Sec:results} reports study results. Section~\ref{Sec:disc} discusses study findings, explores the implications for researchers and practitioners, and states the limitations. Section~\ref{Sec:conc} concludes the paper. 

\section{Related work}
\label{Sec:back}
 \noindent A literature review was done to search for related work. That follows some guidelines of ``Systematic literature studies using snowballing'' from Wohlin et al. \cite{wohlin2014guidelines} but not all. In particular, the guidelines that were followed are the usage of search strings, dedicated search places, explicit inclusion criteria, and documented process. To search for related work, we applied search strings ``software AND test* AND automat* AND (maturity OR mature OR improv*) AND quality AND (release OR effort)" and “software AND test* AND automat* AND (maturity OR mature OR improv*) AND continuous AND integration” in Google Scholar. Though each string got above thousands of search results in Google Scholar, the relevant studies were only presented in the first few pages. The light reading was performed on relevant studies presented in Google Scholar. We did not identify studies working on our research question but found a set of studies (\cite{wang2020,hilton2017trade,khomh2012faster,williams2009,kasurinen2010software,berner2005}) that have observed the relationships between our observed variables (test automation maturity, product quality, test automation effort, and release time) in software development. This set of studies meet inclusion criteria in Table \ref{tab:selectionCriteria}. Next, the Forward snowballing approach was used to screen additional studies based on citations of that set of studies. This screened additional 743 studies in total. The first and third authors read the title, abstract, introduction, and conclusion of each study against inclusion criteria (Table \ref{tab:selectionCriteria}). They used ‘yes’ or ‘no’ to vote for whether a source meets each inclusion criterion. Only 7 studies got ‘yes’ for all inclusion criteria from each author were selected. In the end, in total, 13 studies were selected as related work. We read the full text of each study to identify relevant findings that depict the relationships between any of our observed variables. A summary of these studies is in Table~\ref{tab:related_work}. Column `Variables' shows which variables are observed in the study; Column `Findings' concludes the relevant findings. Next, we provide an overview of these study findings. 
 
\begin{table}[h]
\centering
\caption{Inclusion criteria\label{tab:selectionCriteria}}
\small
\begin{tabular}{ p{15cm} }
\hline
	C1.The study is written in English. \\
	C2.The study is full-text accessible.\\
	C3.The study is not a duplication of others.\\
	C4.The study is published in journals, conferences, workshops. \\
	C5.The study presents findings to depict the relationships between test automation maturity, product quality, test automation effort, and release time in software development. \\
\hline
\end{tabular}
\end{table}
\pagebreak 

{\scriptsize
\begin{longtable}{|p{0.6cm} | p{0.5cm} |p{2.8cm} | p{2.8cm} |p{2.7cm}| p{4.39cm}| }
\caption{Related work}\label{tab:related_work}
\\\hline
	Study & Year & Observation & Methodology & Variables  & Findings \\ \hline
	\cite{wang2020} & 2020 & 
	A test automation maturity improvement program in a DevOps team from a Finnish software company 
	& Experience report (based on meetings, a collection of experience notes, team reflection reports, and telemetry result reports). & Test automaton maturity; release cycle; product quality  & The observed team improved its level of test automation maturity in the CI context using several standard best practices: improve product testability, hire/train expert team members, select and integrate suitable tools, promote communication and collaboration, encourage team members. This team reported that, the higher level of test automation maturity is associated with shorter release cycles and better product quality in the CI context.   \\ \hline

	\cite{lin2020test} & 2020 & 
	148 developers of open-source android apps & Practitioner survey & 
	Test automation maturity; product quality; test automation effort & 91\% of the developers (134/148) confirmed that test automation maturity contributes to app quality, but many reported increased efforts to improve the level of test automation maturity for the promise of high quality apps.    \\\hline
	
	\cite{kumar2016impacts} & 2016 & Three software systems & 
	Case study (based on quantitative observations on metric data from these three software systems) & Test automation maturity; Product quality; Release cycle; Test automation effort & 
	The observations showed that, in these three software systems, test automation maturity have a positive effect on improving product quality, shorting release time, and reducing test automation effort.  \\\hline
	
	\cite{puri2015test} & 2015 & 
	A test automation project in CI context of a communication system & Case study & Test automation maturity; Product quality; Test automation effort; Release cycle & 
	In the observed project, test automation practices were improved against the industrial standards. The result was the increased product quality and reduced test automation effort and release time. \\\hline
	
	\cite{hilton2017trade} & 2015 & 16 software developers from 14 different organizations & Interviews & Test automation maturity; product quality; test automation effort & In the practical CI context, more mature test automation practices can cause better product quality at a cost of increased effort to design rigorous automated tests and execute them in frequent builds. \\ \hline
	
	\cite{ramler2014} & 2014 & A company's programmable logic controller software for machinery &  Experience report & Test automation maturity; test automation effort & When testing this  software, the overall effort invested in adopting disciplined practices to develop rigorous automated tests (reflected as test automation maturity) was high. \\\hline 
	
	\cite{lee2012software} & 2012 & Test maturity improvement program at a small enterprise for developing an medical information system  & Experience report & Test automation maturity; Product quality; Test automation effort & The experience showed that, when test automation practices were improved according to ISO/IEC 29119 (consists of standard practices and techniques), the number of defects after releases (measured for product quality) was largely reduced, though much effort was spent to standardize test automation practices (incl. design, execution, planning).\\ \hline     
	
	\cite{collins2012} & 2012 & A software project developed in the CI context & Experience report & Test automation maturity; Product quality; Release cycle & The project had 20 sprints. Within the project, defects found per sprint were decreased and release speed was increased in the CI context, after improving the level of test automation maturity.  \\\hline
	
	\cite{khomh2012faster} & 2012 & The software development process of Mozilla Firefox during a period when it transitioned to short release cycles
	& Case study (based on data mined from project wiki, project repository, crash repository, bug repository). 
	& Release cycle; Product quality 
	& In the case of Mozilla Firefox, there is no significant association between release cycle frequency and product quality.  \\\hline

	\cite{tosun2009} & 2009 & A software quality improvement project within a health care company in Turkey & Experience report & Test automation maturity; product quality; test automation effort & In the observed project, process maturity was improved against best practices (include some test automation best practices) defined in CMMI (a software process maturity model), and the result was the decrease of 4.5\% defect rate and 17 \% of testing effort. \\\hline

    \cite{williams2009} & 2009 
    & One Microsoft team consisting of 32 developers & Case study based on software repository mining, survey, interview, and action research 
    & Test automation maturity; Product quality; Test automation effort; Release cycle
    & In the observed team, software product quality improved at a cost of approximately 30\% more development time (that delayed release time), when test automation was incrementally performed with disciplined practices under the test-driven development context. \\ \hline

    \cite{kasurinen2010software} & 2009  & 55 testing specialists from 31 organizations & Interviews & Test automation maturity; test automation effort; Release cycle & Test automation practices require considerable effort (in terms of planning effort, design effort, execution effort, and test result analysis effort) to be mature. That may delay release time.\\ \hline
    
    \cite{berner2005} & 2005 & 5 practical test automation projects the authors have participated & Experience report & Test automation maturity; release cycle; product quality & Test automation maturity achieved by using several standard best practices (define a good test automation strategy, carefully design and reuse automated tests, and build a testable SUT) can lead to shorter release cycles and better product quality. \\ \hline
    \end{longtable}}

As noted in Table \ref{tab:related_work}, in all studies \cite{wang2020,lin2020test,kumar2016impacts,puri2015test,hilton2017trade,lee2012software,collins2012,tosun2009,williams2009,berner2005}  that have observed how test automation maturity affects product quality, the findings verified the positive impact. These studies view product quality as to which extent a software product free from defects during the quality assurance process or in the production environment. It is different from scholars who defined product quality as conformance to a standard or to which extent a software product bears on its ability to satisfy given needs~\cite{ISOstandard,garcia2015}. Studies \cite{wang2020,hilton2017trade,lee2012software,berner2005,puri2015test,tosun2009} reported that better product quality can be achieved after improving test automation maturity against certain standard best practices.


The relationship between test automation maturity, product quality, and test automation effort has been discussed from two different viewpoints. One school of viewpoint states the negative relationship of test automation maturity and product quality with test automation effort.  Studies~\cite{lin2020test,hilton2017trade,lee2012software,williams2009,kasurinen2010software}, which hold this view, have observed the increased effort to develop and execute more rigorous automated tests in attaining higher product quality. Experience study \cite{ramler2014} reported that the direct effect of test automation maturity on test automation effort is negative. However, a different school of viewpoint reported by studies \cite{kumar2016impacts,puri2015test,tosun2009} views that, test automation effort will be reduced after improving the level of test automation maturity and product quality after a certain period of time - even though the considerable effort is required to develop and execute rigorous automated tests in attaining high product quality, the later effort savings may offset the invested effort.

The relationship between test automation maturity, product quality, and release cycle also has been viewed differently in prior studies. One view considers that release time must increase against improvements in the level of test automation maturity and product quality. Studies \cite{kasurinen2010software,williams2009} respectively validated this view with qualitative evidence (by interviewing test professionals) and quantitative evidence mined from a software project. This view supports the school of viewpoint that states the negative relationship of test automation maturity and product quality with test automation effort, as noted in the above paragraph. That is, developing and executing rigorous automated tests to ensure high product quality requires considerable effort, so that the elapsed time is added for product development and thus product releases are delayed. In contrast, an alternative view is, the high level of test automation maturity and product quality is accompanied by short release cycles. Studies~\cite{wang2020,collins2012,berner2005} reported this view based upon the authors' experience, while \cite{kumar2016impacts,puri2015test} validated the view in software projects.

Based on the above, prior study findings are not sufficient to answer our research question for many reasons. First, prior studies have declared different viewpoints for the relationships among test automation maturity, product quality, test automation effort, and release cycle. There is a need to evaluate prior viewpoints and solve the conflicts in these viewpoints in CI context of open source projects. Second, only four studies clearly indicated that they focused on the CI context. Third, most prior studies (10/13) reported the experience of individual practitioners or the single case of software organizations/products/projects. To evaluate the viewpoints and draw an overall picture of the software industry, cross-site quantitative evidence is needed~\cite{seaman1999qualitative}. Last, our research aims to validate whether standard best practices in the literature should be recommended for CI practitioners. Among all identified studies, only studies \cite{wang2020,lee2012software,hilton2017trade} had the similar aim. Their findings were based on authors' experience. Many standard  test automation best practices in the literature were not included, e.g., prioritizing important automated tests for execution, providing enough resources, and using the right test automation metrics to measure test automation performance~\cite{wang2019}. To sum up, our empirical study can complement prior studies and make the novelty from the following aspects: 

\begin{itemize}
\setlength\itemsep{0em}
    \item Our study is the first attempt to empirically evaluate the relationships between test automation maturity, product quality, test automation effort, and release cycle in the CI context of open source projects with quantitative evidence.

    \item Our study observed open source projects by surveying their main contributors and mining their repositories (GitHub repositories, CI repositories, and issue tracking systems). Software repository mining methodology was barely used in prior studies. Only study~\cite{williams2009} and \cite{khomh2012faster} have used similar methodology, however, they only mined data from the repositories of a single team or software product, see Table~\ref{tab:related_work}.

    \item Our study was recently finished and intended to present the current view for this research scope. As shown in Table~\ref{tab:related_work}, most of the prior studies were published several years ago, while Only \cite{wang2020,lin2020test} are recent studies.
    
    \item Our findings evidenced some observations of prior studies, resolved the different viewpoints of prior studies, and identified novel research topics to extend the impact and scope of CI and test automation literature. 
\end{itemize}

\section{Research method}
\label{Sec:meth}
\noindent Our research process has five stages: Conceptual model building, Project selection,  Survey design and execution, Project data collection, Data analysis. Each stage is described below.  

\subsection{Conceptual model building} \label{sec:conceptualModel}
\begin{figure}[h]
\centering\includegraphics[width=0.7\linewidth]{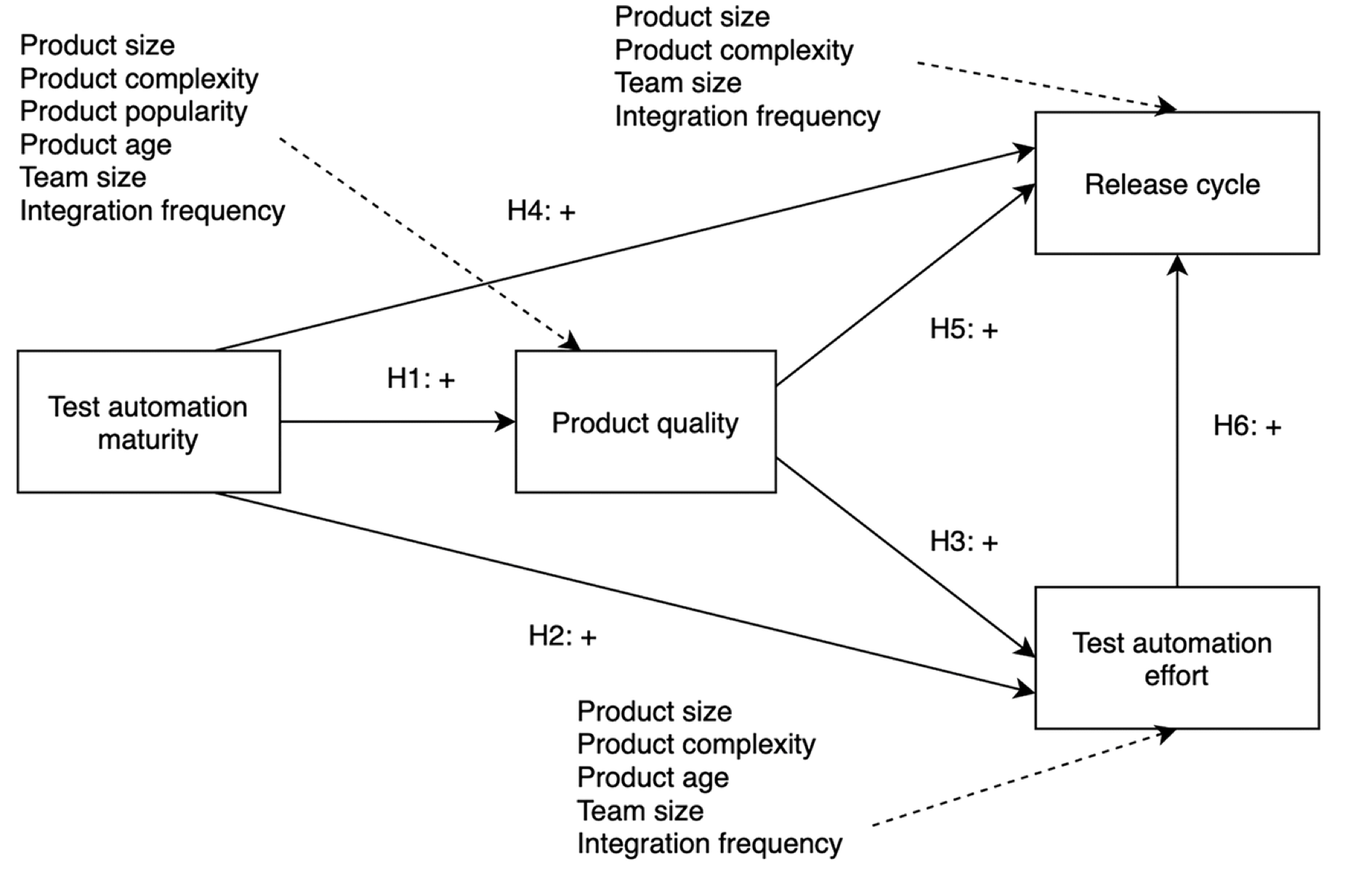}
\caption{Conceptual model}
\label{fig:conceptual_model}
\end{figure}

\noindent 
To address our research question, we built a conceptual model (Figure~\ref{fig:conceptual_model}) to hypothesize the relationships between test automation maturity, product quality, test automation effort, and release cycle. Harter et al. \cite{harter2000} have empirically observed the relationships between software process maturity, product quality, development effort, and cycle time in software development and accordingly proposed a conceptual model. As test automation is a type of software process, to build our model, we adapted Harter et al.’s model by mapping “Process maturity” onto “Test automation maturity” and mapping “Development effort” onto “Test automation effort”. Then, our model is like Harter et al.’s model but with the particular focus on test automation maturity as a type of software process maturity that should have a positive effect on product quality. We also ensured that our model is compatible with previous test automation literature and software engineering literature. The detailed explanation for that is presented in the following sub-sections.

\subsubsection{Observed variables} 
\noindent Our conceptual model contains four observed variables. These four variables were derived from our research question and they were conceptualized as below:

\begin{itemize}
\setlength\itemsep{-0.1em}
    \item \textbf{Test automation maturity:} the level of test automation maturity assessed by standard best practices in the literature. That is, to which extent test automation practices within an organization are close to standard test automation best practices defined in the literature. 
    
     \item \textbf{Test automation effort:} the total effort to develop and execute automated tests.
     
     \item  \textbf{Release cycle:} the length of release cycles - the average time it takes to release a product from the start of development \cite{harter2000}. 
     
     \item \textbf{Product quality:} 
     our conceptual model utilizes the common definition of product quality as prior studies in Table \ref{tab:related_work}. It considers product quality as to which extent a software product free from defects in the production environment.
    
\end{itemize}

\subsubsection{Hypothesis} 
\noindent Our conceptual model makes six hypotheses grouped into three aspects: 

\paragraph{Effect of test automation maturity on product quality} Harter et al.'s model observed that, high levels of software process maturity (assessed by standard best practices) aid in detecting early defects and reducing defect injection, and thus, are accompanied by high product quality. In test automation literature, prior studies \cite{wang2020,puri2015test,hilton2017trade} (Table \ref{tab:related_work}) also reported that, better product quality can be achieved using certain standard best practices for test automation maturity improvement in the CI context. This implies the hypothesis:

\begin{itemize}
    \item \textbf{H1 (Test automation maturity and Product quality).} \HoneContent.
\end{itemize}

\paragraph{Effect of test automation maturity and product quality on test automation effort} Harter et al.’s model observed that, increased development effort was required to follow standard best practices to achieve higher levels of software process maturity and product quality. As discussed in Section \ref{Sec:back}, in test automation literature, the relationship between test automation maturity, product quality, and test automation effort has been discussed from two different viewpoints. To be compatible with Harter et al.’s observation, our conceptual model follows the view, that implies, there is an inverse relationship of test automation maturity and product quality with test automation effort. Thus, two hypotheses are built: 

\begin{itemize}
\setlength\itemsep{0em}
    \item \textbf{H2 (Test automation maturity and Test automation effort)}. \HtwoContent.
    
    \item \textbf{H3 (Product quality and Test automation effort).} \HthreeContent. 
\end{itemize}

 \paragraph{Effect of test automation maturity and product quality on release cycle} Harter et al.’s model observed that, increased development
efforts invested in achieving higher levels of software process maturity and product quality would lead to the delay of cycle time. As noted in Section \ref{Sec:back}, in test automation literature, the relationship between test automation maturity, product quality, and release cycle also has been viewed from two different viewpoints. To be compatible with Harter et al.’s observation, our model follows the view, which indicates cycle time will increase against improvements in the level of test automation maturity and product quality. This supposes:  
 
 \begin{itemize}
 \setlength\itemsep{0em}
     \item \textbf{H4 (Test automation maturity and Release cycle).} \HfourContent.
     
     \item \textbf{H5 (Product quality and Release cycle).} \HfiveContent.
     
     \item \textbf{H6 (Test automation effort and Release cycle).} \HsixContent.
     
 \end{itemize}

\subsubsection{Control variables}
\noindent Scientists use control variables to ensure their experimental results are solely caused by their observations \cite{field2002design}. In our conceptual model, we identiﬁed control variables that may affect our observations on the relationships among our observed variables. Each control variable is explained below:

\begin{itemize}
\setlength\itemsep{0em}
    \item \textbf{Product size:} how many lines of code (LOC) a software product contains \cite{harter2000}.

    \item \textbf{Product complexity:}  ``software complexity is a natural byproduct of the functional complexity that the code is attempting to enable" \cite{samli2020}.
    
    \item \textbf{Product popularity:} how many forks a product has \cite{vasilescu2015}.
    
    \item \textbf{Product age:} software product age is the time a software product is created. 
    
    \item \textbf{Team size:} how many people are work in software development \cite{vasilescu2015}.
    
    \item  \textbf{Integration frequency:} how often the code changes are integrated in CI context~\cite{staahl2017continuity}.  
\end{itemize}

\paragraph{Product quality} Studies \cite{banker1992software,agrawal2007,zazworka2011} empirically observed that products in larger size have more defects. Many scholars \cite{agrawal2007,zazworka2011,maccormack2006} observed a positive correlation between product complexity and product quality: more complex software products tend to have more defects. Harter et al. \cite{harter2000} investigated that newer products expose more defects. A recent study~\cite{vasilescu2015} examined that products that are widely forked may have a high opportunity to expose defects. Many studies \cite{cotroneo2016bugs,ibrahim2012relationship,au2009virtual} reported that software products developed in a larger team size tend to have more defects, because ``defects are human being caused'' and thus more people means a higher occurrence chance of defects. Studies \cite{staahl2017continuity,vasilescu2015} examined that product quality is affected by integration frequency.

\paragraph{Test automation effort} Based on many studies \cite{staahl2017continuity,harter2000,agrawal2007}, product size is an important factor that explains test automation effort, as test scope and frequency tend to increase along with product size growth. Several scholars \cite{harter2000,agrawal2007} empirically observed that it would take more effort to develop and execute automated tests on software products with higher complexity. As noted by the study \cite{kumar2016impacts}, older products spend less test automation effort as they can reuse the existing automated tests. Moreover, many studies \cite{staahl2017continuity,banker1992software,zhao2017impact,shahin2017continuous} have noted that larger teams spend more test automation effort in the CI context. More developers in the team indicate an increased rate of changes - to make a software product to be ready on each committed change, test automation should reach the proper scope and deep - thus test automation effort is increased  \cite{staahl2017continuity,banker1992software,zhao2017impact,shahin2017continuous}. Prior scholars \cite{hilton2016usage,hamdan2015quality} validated that integrating code changes cost test automation effort in CI builds.

\paragraph{Release cycle} Studies \cite{harter2000,agrawal2007} found that product size and product complexity are  negatively associated with release time. Larger products take longer to be developed and therefore are accompanied by longer release cycles \cite{harter2000,agrawal2007}. More complex products are delivered in longer release cycles, because the longer time is required to code more complex functions and methods inside of them \cite{harter2000,agrawal2007}. The study \cite{vasilescu2015} evidenced that the team size could affect software development productivity that determines the release time in the CI context. Studies \cite{hamdan2015quality,staahl2017continuity,pinto2018work} observed that frequent integration would lead to short release cycles.

\subsection{Project selection}
\label{sec:project_selection}

\noindent To validate our conceptual model (Figure \ref{fig:conceptual_model}) in the CI context of open source projects, it was decided to select  open source java projects that adopt test automation in the CI context. As the performance and structure of different programming languages are different, to avoid bias of the research, we must select projects written in the same programming language. Java projects were focused since they are popular and exist in large numbers; This would enable us to find enough relevant projects for our study. We began with JTeC \cite{coro2020jtec} and 20-MAD~\cite{claes202020} datasets that have collected the amount of open source java projects. The project selection process consists of two steps. Table \ref{tab:selectionResults} summarizes each step's result.

\begin{table}[h]
\small
\centering
\caption{Summary of project selection results}\label{tab:selectionResults}
\begin{tabular}{ llll }
\hline
Dataset & Original & Step 1 & Step 2 \\
\hline
	JTec \cite{coro2020jtec} & 31222 & 1459 & 126 \\ \hline
	20-MAD  \cite{claes202020} & 765 & 38 & 23 \\ \hline
\end{tabular}
\end{table}

In step 1, we screened projects from JTeC and 20-MAD datasets against the initial selection criteria in Table \ref{tab:project_selection_criteria}. Projects that did not meet all criteria were excluded. C1 ensured that a project is original. C2 checked whether a project adopts CI and its CI repository is accessible to us. We only looked at projects that use Travis CI. Travis CI is a popular CI server tool for open source projects and it allows to access the building history via its API. That would allow us to collect test automation and CI related data for our study.  C3 verified whether a project has run automated tests in its CI context, and limited the building tool to Maven. As the performance of different CI building tools to compile and run automated tests is different \cite{hilton2016usage}, to avoid bias of the research, it is necessary to focus on one CI building tool. Maven has a standard output for test execution and allows externals to mine testing related data. It is dominant in Apache open source java projects\footnote{https://github.com/apache}, which exist in a large number on our starting datasets. Ignoring Apache projects would lose the large number of candidate projects. C4 selected the projects that report bugs using Github issue tracker or Jira, which are common issue tracking systems for java projects. At end of step~1,  the initial selection criteria resulted in a candidate set of 1,497 projects.  

\begin{table}[H]
\small
\centering
\caption{Initial selection criteria}\label{tab:project_selection_criteria}

\begin{tabular}{ p{13cm} }
\hline
	
	C1. Is not a fork repository of the other.\\
	
	C2. Use Travis CI and its Travis CI repository is accessible to the public.\\
	
	C3. Run automated tests with Maven under Travis CI environment.
	
	C4. Use Github issue tracker or Jira. \\
\hline
\end{tabular}
\end{table}

In step 2, we further screened the candidate projects selected from step~1 against project active selection criteria in Table \ref{tab:active_selection_criteria}.
These criteria were used to select projects, which are active to commit changes~(AC1), run CI builds (AC2), adopt test automation with our selected CI tool Maven (AC3), report defects in issue tracker system (AC4-6), and publish releases (AC7) at the time of doing our research. At the start, we examined prior software repository mining studies~\cite{hilton2016usage,vasilescu2015,rausch2017,cataldo2009,subramanian2007} on our observed variables to set threshold values in these criteria. Many attempts were conducted to test threshold values, e.g., changing from a dozen to thousands of commits, CI builds, automated tests, reported defect-related issues, and the number of releases. The final set of threshold values was chosen according to the results of our attempts, as it allowed us to select projects (from datasets) that contain enough metric data on our observed variables and ensure sample diversity~\cite{linaaker2015guidelines} for valid observations. Our attempts found that, most projects that did not meet AC1 are inactive - their project repositories were barely updated or changed in the current year. Collecting data from such projects may make invalid observations for their current test automation practices. Regarding AC2, our attempts showed that, projects did not meet AC2 and AC3 having few automated tests and these projects exist in large numbers. Including these projects may introduce too many outliers \cite{miller1993tutorial}. That may have the disproportionate effect on later statistical results. We also examined that, setting the threshold value of CI builds to bigger numbers (e.g., 20, 50, 100) in AC2 are likely to neglect projects, which are in small size or just start CI and test automation. Neglecting such projects may harm the sample diversity. Our attempts showed that, most of the projects that did not meet AC4-AC6 were not active in issue tracker systems - we were only able to recognize 0-2 bugs from the issue tracker system of each of such projects. Including these projects may bias our research, which intended to measure product quality of projects using reported bugs from issue tracker systems. AC7 was set to select projects with varied release cycles but reject the null value for statistics. As a result, we finally selected 149 projects that met all criteria for further study.

\begin{table}[H]
\small
\centering
\caption{Project active selection criteria}\label{tab:active_selection_criteria}

\begin{tabular}{ p{15cm} }
\hline
	AC1. Had at least 90 commits in 2020\\
	
	AC2. Run at least 10 CI builds under Travis CI environment in 2020 \\
    
    AC3. Run automated tests with Maven in CI builds in 2020 \\
    
	AC4. Had more than 90 issues reported in Github Issue Tracker or Jira
	
	AC5. Used tags to label issues - at least 60\% of issues have tags
	
	AC6. Used naming convention (e.g.,  bug,  defect,  fault,  error, flaw, and other synonyms) to label defects in Github Issue Tracker or Jira\\
	
	AC7. Had at least a release in 2020  \\
\hline
\end{tabular}
\end{table}

\subsection{Survey design and execution} \label{sec:surveyDesign}
\noindent To explore the state of practice of test automation maturity (regarding the adoption of standard best practices in the literature) in selected 149  projects, we conducted a test automation maturity survey with the main contributors of each project. For survey design and execution, authors consulted ‘Guidelines for Conducting surveys in Software Engineering’ from Linåker et al. \cite{linaaker2015guidelines} and the general survey guidelines from Groves et al.~\cite{groves2011survey}.

\paragraph{\textbf{Sampling plan}} To ensure the accuracy of the sample, the first author created a sampling plan and it was reviewed and revised with other authors. In survey studies, two common sampling methods are widely used: probabilistic sampling and non-probabilistic sampling~\cite{punter2003,groves2011survey}. Probabilistic sampling assumes that every member of the target population is available and there is a random process to select participants \cite{groves2011survey}. Non-probabilistic sampling requires selecting the representative sample \cite{groves2011survey}.

In our survey, the target population was defined as the main contributors of selected 149 projects. Non-probabilistic sampling method was used for two reasons. First, due to European General Data Protection Regulation (GDPR), it was impossible to collect all individual contributors' contacts and send out the survey to them without their permission. Second, open source projects allow for the involvement of both internal and external contributors. In our survey, it was expected to only select internal contributors, who currently are core members of a project and really understand test automation practices and CI practices in this project. As a consequence, for each of 149 projects, we visited its project page and browsed the list of internal contributions, and then, only selected contributors meeting the following criteria:

\begin{itemize}
\setlength\itemsep{0em}
    \item Currently is the core member of the project.
    \item Had more than 10 commits to the project in recent 6 months.
    \item Choose to make their email addresses visible to the public. This criterion was defined with respect to GDPR. 
\end{itemize}

 Each project had 3-77 contributors meeting all of the above criteria. In total, 1432 contributors from 149 projects were selected as our sample to distribute our test automation maturity survey. 

\paragraph{\textbf{Survey design}} 
We reused our previous test automation maturity survey developed in our 2020 study~\cite{icsoft20}. That survey was developed using a knowledge base  established upon 18 test maturity models \cite{icsoft20,wang2019}. It has been reviewed by test automation experts and executed with 151 practitioners (coming from above 101 organizations in 25 countries) in the industry. We tailored that survey to fit the needs of our study in this paper. Our survey in this paper contains three parts.

Part 1 contains a consent form, which introduces the survey content, shows the information of principal researchers and organizations who designed the survey, and declaims the research purpose. Only respondents, who consent that they are the right audience of the survey and are willing to participate in the research, are directed to Part 2, otherwise, the survey is closed.

Part 2 presents 16 test automation maturity questions (Table \ref{tab:survey_part2}). These maturity questions state standard best practices for different key areas of test automation. We designed that respondents answer each maturity question using an ordinal scale: 1- strongly disagree, 2 - disagree, 3 - slightly disagree,
4 - slightly agree, 5 - agree, 6 - strongly agree. The higher agreement selected in the scale reveals that test automation practices carried out in respondents' organizations are more close to the standard best practice stated in a maturity question. Besides, ‘no answer’ options were provided.  Respondents were allowed to leave comments in a free-text field situated at the end.

\begin{table}[ht]
  \begin{center}
   \footnotesize
    \caption{Test automation  maturity questions \cite{icsoft20}}
    \label{tab:survey_part2}
    \begin{tabular}{p{16cm} }
     \hline

      \textbf{SQ1-Strategy}. We have a test automation strategy that defines `what test scope will be automated to what degree, when, by whom, by which methods, by what test tools, in what kind of environment'.\\ 
      \hline
      \textbf{SQ2-Resources}. We allocate enough resources for test automation, e.g., skilled people, the funding, the time \& effort, test environment with the required software, hardware, or test data for test automation.\\\hline
      
      \textbf{SQ3-Roles}. We clearly define roles and responsibilities of stakeholders in test automation.  \\\hline
      
      \textbf{SQ4-Knowledge}. We are systematically learning from prior projects. We collect and share expertise, good test automation practices, and good test tools for future projects.  \\\hline
      
     \textbf{SQ5-Competence}. Our test team has enough expertise and technical skills to build test automation based on our requirements.  \\\hline
     
      \textbf{SQ6-Tools}. We currently have the right test tools that best suit our needs.  \\\hline
      
      \textbf{SQ7-Test environment}. We have control over the configuration of our test environment. \\\hline
      
      \textbf{SQ8-Guidelines}. We have guidelines on designing and executing automated tests. Those guidelines include, e.g., coding standards, test-data handling methods, specific test design techniques to create test cases, processes for reporting and storing test results, the general rules for test tool usage, or information on how to access external resources.  \\\hline
      
      \textbf{SQ9-Prioritization.} We effectively prioritize and schedule automated tests for the execution.  \\\hline
      
      \textbf{SQ10-Results.} We are capable to manage and integrate test results collected from different sources (e.g., different test tools, test levels, test phases) into a big picture, and then report useful information to the relevant stakeholders.  \\\hline
      
      \textbf{SQ11-Process.} We organize our test automation activities in the stable and controllable test process.\\\hline
      
       \textbf{SQ12-SUT.} Our Software Under Test enables us to conduct our test automation, e.g., maturity, running speed, or testability of our Software Under Test is not a problem for our test automation.\\\hline
      
      \textbf{SQ13-Measurement.} We have the right metrics to measure and improve our test automation process.\\\hline
      
      \textbf{SQ14-Testware}. Our testware (e.g., test cases, test data, test results, test reports, expected outcomes, and other artifacts generated for automated tests) is well organized in a good architecture and it is easy to be maintained. \\\hline
      
     \textbf{SQ15-Efficient\&Effective.} We create automated tests that are able to produce accurate and reliable results in timely fashion. \\\hline
     
     \textbf{SQ16-Satisfaction.} We create automated tests can meet the given test purposes and consequently bring substantial benefits for us, e.g., better detection of defects, increase test coverage, reduce test cycles, good Return on Investment, better guarantee product quality. \\
\hline
    \end{tabular}
  \end{center}
  \end{table}

Part 3 presents several background questions to the project and respondent profile. The purpose is to check the background of respondents and understand in which situations test automation practices are carried out in the CI context of selected projects.

\paragraph{\textbf{Survey distribution}} 
Our survey was hosted by an online survey tool called LimeSurvey\footnote{https://www.limesurvey.org/} and it was available to invitees (1432 contributors from 149 projects) from 18th October 2020 to 18th January 2021.  We sent out personalized email invitations to invitees on 18th October 2020. Later, two rounds of reminder emails were sent out respectably in November and December 2020. Invitees were asked to complete our survey online. The participation was voluntary and anonymous. The survey and its host tool LimeSurvey adhere to GDPR. Invitees were allowed to withdraw their responses at any time. To attract invitees, we established reward mechanisms. When our survey finished, we have sent an individual report that summarizes a snapshot of all survey responses to each respondent. We selected five lucky respondents and gave each a 50-euro Amazon gift card.

In total, we received 43 responses from 41 projects to our survey.  The response rate counted for project level is 27.5\% (41 out of 149 projects). Since our study in this paper aimed to make observations on the project level, the response rate of our survey also is counted at the project level.

\paragraph{\textbf{Response quality control}} To control the quality of our survey responses, we consulted the online survey standard from Ganassali \cite{ganassali2008}. To improve the overall quality of our survey, we removed six survey responses from our pool. These six responses either had the same answer on over half of all maturity questions, or had selected 
`no answers’ on over five maturity questions. As such, a final pool of 37 responses from 37 projects was finally built for our study in this paper.

\paragraph{\textbf{Project and respondent profile}} 
Table \ref{tab:TA_situations} depicts the project and respondent profile of 37 projects in our final pool. We see that test automation coverage is rather high in these projects. Approximately 75\% of the projects automated over 50\% of test cases. 37.8\% of the projects automated at least 90\% of test cases. All projects automated unit tests and above half of the projects automated integration tests, while automated tests in other levels take a relatively small proportion in these projects. All respondents are internal contributors and most of them are developers (78.4\%) in the projects. Many respondents (73\%) have been working on test automation in CI context of these projects for 1-5 years.

\begin{table}[ht]
\small
\centering
\caption{Project and respondent profile}\label{tab:TA_situations}
\begin{tabular}{ cc }  
\begin{tabular}{ |l l|} 
\hline
     & Response \\\hline
	 \textbf{\% of automated test cases:} &  \\ 
	 $<$10\% & 2 (5.4\%)\\
	 11-50\% & 4 (10.8\%)\\
	 51-90\% & 14 (37.8\%)\\
	 $>$90\% & 14 (37.8\%) \\ 
     We don't measure & 3 (8.1\%)\\\hline 
     
     \textbf{Test-level automation$^a$:} & \\ 
     Unit & 37 (100\%)\\
     Integration & 19 (52.8\%) \\
     System & 6 (16.7\%)\\
     Acceptance & 2 (5.4\%) \\
     Performance & 3 (8.1\%) \\
     Regression & 13 (36.1\%)\\
     GUI & 7 (19.4\%)\\
     Stress & 1 (2.8 \%)\\ \hline
     
    \multicolumn{2}{l}{\footnotesize $^a$ This is a multiple choice question.} \\

\end{tabular} &  
\begin{tabular}{ |p{5cm} l| } 
\hline
& Response \\\hline
	\textbf{Current role in projects:} &  \\
	Test Lead/Manager/Director & 5 (13.5\%)\\
    Developers & 29 (78.4\%) \\
    Testers & 3 (8.1\%)\\\hline
    
    \textbf{Years of working test automation in CI context of the projects:} &  \\ 
    6-10 years & 7 (18.9\%)\\
    1-5 years & 27 (73.0\%) \\
    Less than 1 year & 3 (8.1\%)\\ 
    
\hline
\end{tabular} \\
\end{tabular}
\end{table}

\paragraph{\textbf{Response overview}}
Figure 2 shows the overview of responses to  maturity questions.  Agreed responses
(slightly agree - strongly agree) are stacked to the right of a vertical baseline on ‘0\%’ on the x-axis. Disagreed responses (slightly disagree - strongly disagree) are stacked to the left of the same baseline. Note that, since ‘no answer’ exists for certain questions, the total percentage of each question may be not equal to 100\%. ``SQ6-Tools" and ``SQ5-Competence'' got the highest percentage of agreed responses (88\% and 86\%), suggesting that, most projects have the right test tools and competent test professionals to conduct test automation in their current CI context. ``SQ9-Prioritization'' and ``SQ10-Results'' have the highest percentage of disagreed responses (51\%), suggesting that, many projects still are not capable to effectively prioritize automated tests and handle test automation results in their current CI context. In our previous study \cite{wang2020} that surveyed 151 practitioners using the same survey, ``SQ5-Competence'' also got the highest percentage of agreed responses (85\%), while ``SQ8-Guidelines" (47\%) got the highest percentage of disagreed responses indicating there is a lack of guidelines on designing and executing automated tests in general. 

As described in the above survey design related paragraphs, accompanying with maturity questions, there is a free-text field situated to allow respondents to leave free comments. However, there was no respondent leave free comments in that field.    

\begin{figure}[H]
\centering\includegraphics[width=1\linewidth]{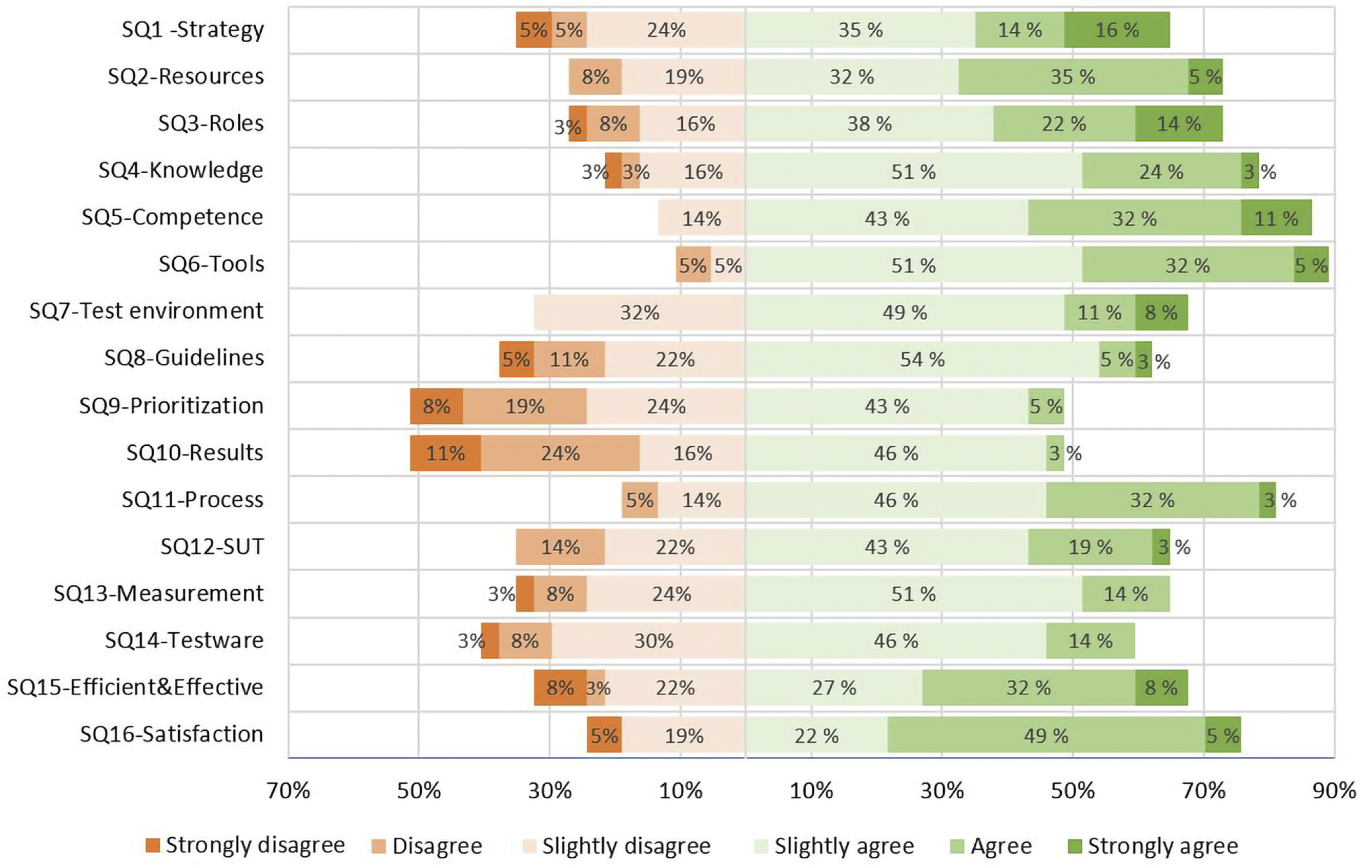}
\caption{Survey response}
\label{fig:survey_result}
\end{figure}

\subsection{Project data collection}
\label{sec:project_data_collection}
\noindent We collected data from 37 projects in our final pool. Two snapshots were set on each project. Snapshot 2 represents the state of the project on 18th January 2021 (the date when our test automation survey closed). Snapshot 1 represents the state of the project on 18th January 2020 (one year before Snapshot 2).  We studied each project with its development practices between these two snapshots. Two snapshots had a one-year interval. The reason to set one-year interval is: short interval based data (e.g., three-month-interval) may be insufficient to provide a stable overview of the project's release cycles, test automation effort, and product quality; long interval based data (e.g., two-year-interval) may not correspond to the project's current state of test automation practices (that we surveyed in our test automation maturity survey). Table \ref{tab:metrics} shows the metrics we used to measure variables in our conceptual model~(Figure \ref{fig:conceptual_model}). To collect metric data, we collect survey responses, and mined each project's GitHub repository, CI repository, and issue track system. Table \ref{tab:sum_static_metics}  presents summary statistics for collected metric data.

\begin{table}[ht]
\small
\centering
\caption{Variable metrics}\label{tab:metrics}
\begin{tabular}{ l   p{11cm}  }\hline
	\textbf{Observed variable} & \textbf{Metric} 
	\\ \hline
	 
	Test automation maturity &  Total score of all maturity questions in part~2 of our test automation maturity survey \\ \hline
	
	Product quality & Defect density = the cumulative number of post-release bugs between two snapshots/  KLOC$^b$ of production at Snapshot 2  \\ \hline
	
	Test automation effort &  
    Development effort (in seconds) = Increased LOC (without comments and blanks) of test automation between two snapshots \newline
    Execution effort (in seconds) = Total time to run all automated tests in CI builds between two snapshots \\ \hline 
    
	Release cycle  & The number of releases between two snapshots \\\hline

   \textbf{Control variable} & \textbf{Metric}  \\ \hline
    Product size & LOC$^c$ (without comments and blanks) of the current product at Snapshot 2  \\ \hline

    Product complexity & 
    Average Cyclomatic complexity number on all coding files of the current product at Snapshot 2 
     \\ \hline
    
    Product age & Product age in days since the date of the first commit to the date of Snapshot 2   \\ \hline
    
    Team size & The average number of contributors per month between two snapshots \\ \hline
    
    Product popularity & The number of forks at Snapshot 2 \\\hline
    
    Integration frequency & The number of CI builds between two snapshots \\\hline

    \multicolumn{2}{l}{\footnotesize $^b$ KLOC:thousands of lines of code.} \\
    \multicolumn{2}{l}{\footnotesize $^c$ LOC:lines of code.} \\
\end{tabular}
\end{table}

\begin{table}[!htbp] 
\small
\centering 
  \caption{Summary statistics on variable related metric data 
  }
  \label{tab:sum_static_metics} 
\begin{tabular}{@{\extracolsep{5pt}} p{4.5cm} lllll} 
\\[-1.8ex] 
\hline \\[-1.8ex] 
Variable  & \multicolumn{1}{l}{Mean} & \multicolumn{1}{l}{Median} & \multicolumn{1}{l}{St. Dev.} & \multicolumn{1}{l}{Min} &  \multicolumn{1}{l}{Max} \\ 
\hline \\[-1.8ex] 

 Test automation maturity & 61.865 & 64 & 12.383 & 30 & 84 \\ \hline 

 Defect density \newline (meas.Product quality)  &  0.566 & 0.399 & 0.699 & 0.037 &  3.737 \\ \hline 

Test automation effort: &  & & & \\ 
Development effort & 6,852.595 & 1614 & 21,120.510 & 17 & 127,113 \\
Execution effort & 2,517,660 & 299341 & 8,079,574 & 3,078 & 48,158,813 \\ 
\hline

Number of releases \newline (meas.Release time) & 8.757 & 6 & 7.429 & 1 & 29 \\ \hline

Product size  & 229,234.200 & 77,166 & 398,647.300 & 3,479 & 1,868,078 \\\hline

Product complexity & 1.997 & 1.8 & 0.639 & 1.200 & 4.600 \\ \hline

Product age & 3,765.378 & 3762 & 1,051.079 & 578 & 6,782 \\ \hline

Team size & 5.135 & 3 & 5.213 & 1 & 27 \\ \hline

Product popularity & 5,212.838 & 398 & 15,225.550 & 10 & 86,000 \\  \hline

Integration frequency & 739.9 & 506 & 777.165 & 28 & 2620\\\hline
\end{tabular} 
\end{table} 

``Test automation maturity'' was measured by the total score of all maturity questions in our survey. For each maturity question, the answer on an ordinary scale was recorded in a score of 1-6 (from strongly disagree to strongly agree); ‘no answer’ converted to ‘0’. The total score sums the score of all maturity questions. The higher total score represents that, test automation practices within a respondent’s organization are more close to standard best practices in the literature suggesting a higher level of test automation maturity.

The metric for ``Product quality'' was ``Defect density'' that is widely used by prior scholars \cite{agrawal2007,harter2000}. Defect density measures the defects related to the product size -  the larger value of defect density imply a higher frequency of defects per unit of a product, i.e., lower product quality. In our study, defect density was computed as the cumulative number of post-release defects (reported by contributors and users) divided by product size at Snapshot~2. Post-release defects were used as they are the ones neglected by CI practices, and thus, can be used to observe how CI practices (with test automation maturity) is effective to improve product quality \cite{rwemalika2019industrial}.  In step 2 of our project selection process, we have selected projects that ``use naming convention  (e.g.,  bug,  defect,  fault,  error, flaw, and other synonyms) to label defects" in their issue tracker system, see Table \ref{tab:active_selection_criteria} in Section \ref{sec:project_selection}. Thus, to compute the cumulative number of post-release defects reported between two snapshots for each project, we count the number of issues assigned with defects-related labels after releases between two snapshots in each project's issue tracker system.

\begin{figure}[h]
\centering\includegraphics[width=0.7\linewidth]{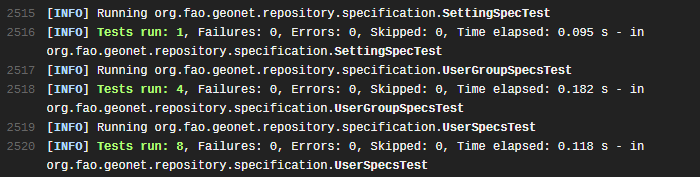}
\caption{The standard output of Maven tests from Core-geonetwork}
\label{fig:mavan_output}
\end{figure}

 In our conceptual model, `Test automation effort'' is the combination of test automation development effort and execution effort. Separate metrics were used for measuring test automation development effort and execution effort. Each project's test automation development effort was measured by increased LOC (without comments  and blanks) of test automation between two snapshots.  To measure software development effort, software engineering researchers have discovered different metrics: LOC, Person-days (the time in days required for one person to complete a task), active-days (the number of days contributors make a commit) \cite{jorgensen1995,shihab2013lines,boehm2000software,bergeron1992}. In this paper, LOC was used as it is more widely used than other metrics for measuring the development effort of open source projects \cite{jorgensen1995,shihab2013lines,boehm2000software,bergeron1992}. Besides, to measure test automation execution effort for each project, the total time to execute all automated tests in CI builds between two snapshots was computed. For each project, we mined test log files to collect the total time of all CI builds (occurred between two snapshots) for executing automated tests. Maven standard output provided a convenient way to do that. Figure \ref{fig:mavan_output} shows an example from the project Core-geonetwork\footnote{https://travis-ci.org/github/geonetwork}, with Maven standard output, the execution time is printed out for units of automated tests. In the end, to combine test automation development effort and execution effort into test automation effort, we did the data normalization, as the scale of execution effort is different from the scale of development effort, see Table \ref{tab:metrics} and \ref{tab:sum_static_metics}. Test automation development effort and execution effort were normalized on a scale of 1-100 separately for each project. The sum of the normalized development effort and execution effort was computed as ``Test automation effort'' for each project. 
 
`Release cycle' was measured by the number of releases between two snapshots - the more number of releases indicates shorter release cycles for a project. 

All control variables were measured at Snapshot 2 that represents the current state of a project. Product size was measured by LOC developed in the current product excluding comments and blanks. The use of LOC is in line with prior scholars \cite{agrawal2007,harter2000}. Average Cyclomatic complexity number \cite{mccabe1976complexity} was computed (with a tool Lizard\footnote{https://github.com/terryyin/lizard}) to determine product complexity. It measures the number of linearly independent paths in source code \cite{mccabe1976complexity}. A lower average Cyclomatic complexity number indicates the lower product complexity. As contributors may join or leave at any phases of open source projects \cite{vasilescu2015}, a project's team size was measured as the average number of contributors per month between two snapshots. Product popularity was measured by the number of forks of a project at Snapshot 2 - a large number of forks indicates higher product popularity. Integration frequency was measured by the number of CI builds between two snapshots - more number of CI builds indicates higher integration frequency.  

\subsection{Data analysis}
\begin{figure}[h]
\centering\includegraphics[width=0.8\linewidth]{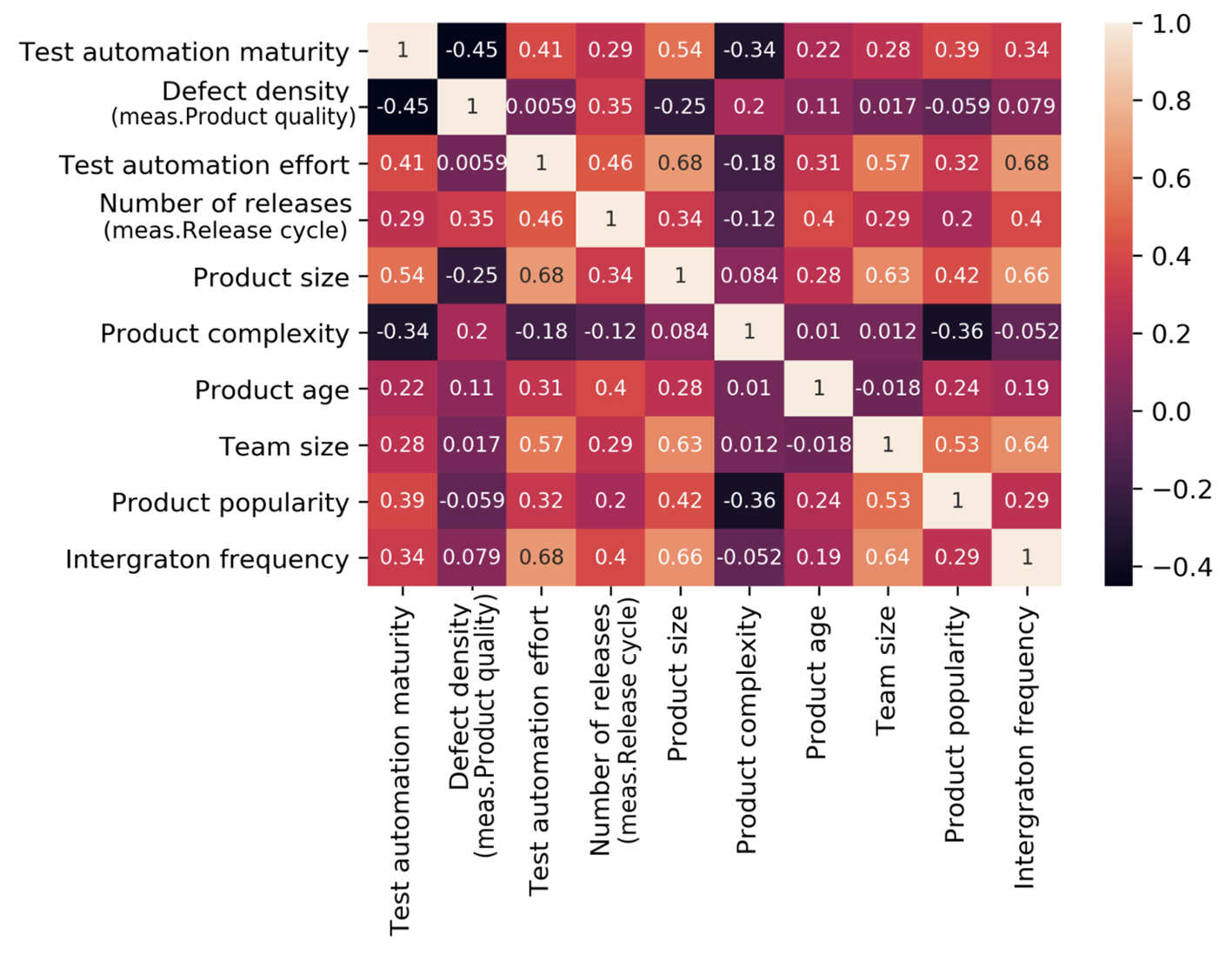}
\caption{Correlation Matrix}
\label{fig:corr_martix}
\end{figure}

\noindent We first observed if there is a correlation between variables in our conceptual model. Figure~\ref{fig:corr_martix} presents the correlation matrix of metric data on our variables. Spearman method was used to compute the correlation matrix, as it can measure the strength of both linear and non-linear association between two variables \cite{croux2010influence}. As uni-variate correlations are not solid statistical evidence, it is imperative to run regression analysis to further study our conceptual model with metric data.

 To validate our conceptual model (Figure \ref{fig:conceptual_model}) that hypothesized the relationships among our observed variables, we developed several empirical models:

\begin{enumerate}
    \item EM1. Product quality = \textit{f}(Test automation maturity, product size, Product complexity, Product popularity, Product age, Team size, Integration frequency) 
    
    \item EM2. Test automation effort = \textit{f}(Test automation maturity, Product quality, Product size, Product complexity, Product age, Team size, Integration frequency) 

    \item EM3. Release cycle = \textit{f}(Test automation maturity, Product quality, Test automation effort, Product size, Product complexity, Team size, Integration frequency) 
\end{enumerate}

Each empirical model addresses at least one hypothesis in our conceptual model. Table \ref{tab:EM_H_corr} shows which hypothesis(es) each empirical model addresses. We applied multiple regression on our metric data  of our variables to study our empirical models. However, before starting regression analysis, we did further observations on metric data of our variables.

\begin{table}[H]
\small
\centering
\caption{Empirical model v.s. hypothesis}
\label{tab:EM_H_corr}
\begin{tabular}{ l p{13cm} } \hline
   Empirical model & Hypothesis \\\hline
   EM1 & H1. \HoneContent. \\\hline
   EM2 & H2. \HtwoContent; \\
       & H3. \HthreeContent. \\\hline
   EM3 & H4. \HfourContent; \\
       & H5. \HfiveContent; \\
       & H6. \HsixContent. \\\hline
\end{tabular}
\end{table}

\begin{figure}[H]
    \centering
    \subfigure[EM1]{\includegraphics[width=0.43\textwidth]{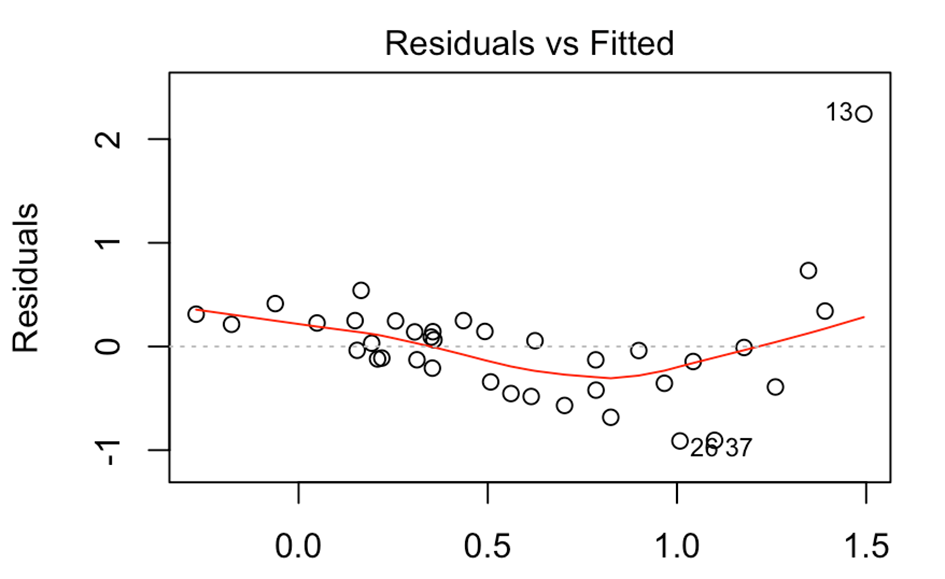}} 
    \subfigure[EM2]{\includegraphics[width=0.43\textwidth]{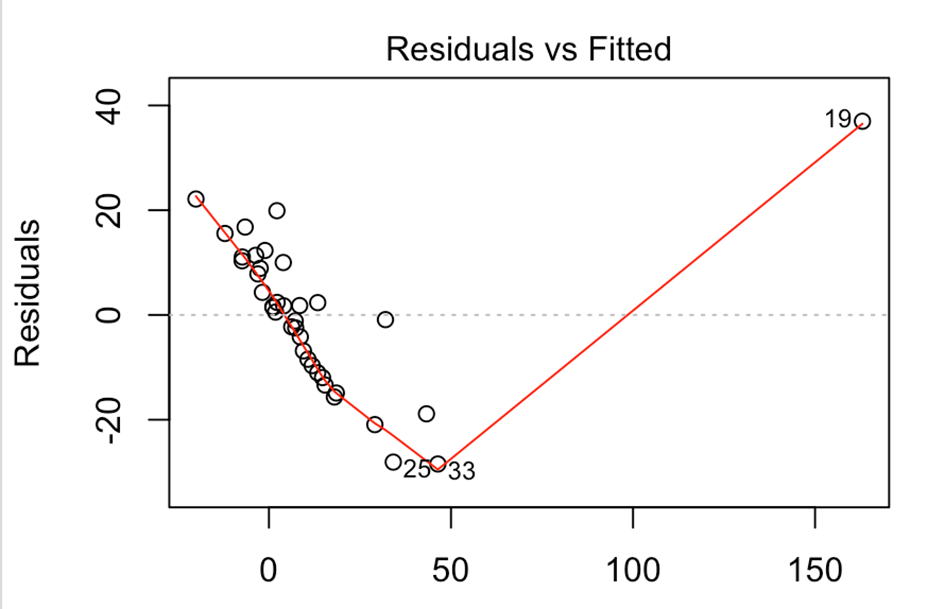}} 
    \subfigure[EM3]{\includegraphics[width=0.43\textwidth]{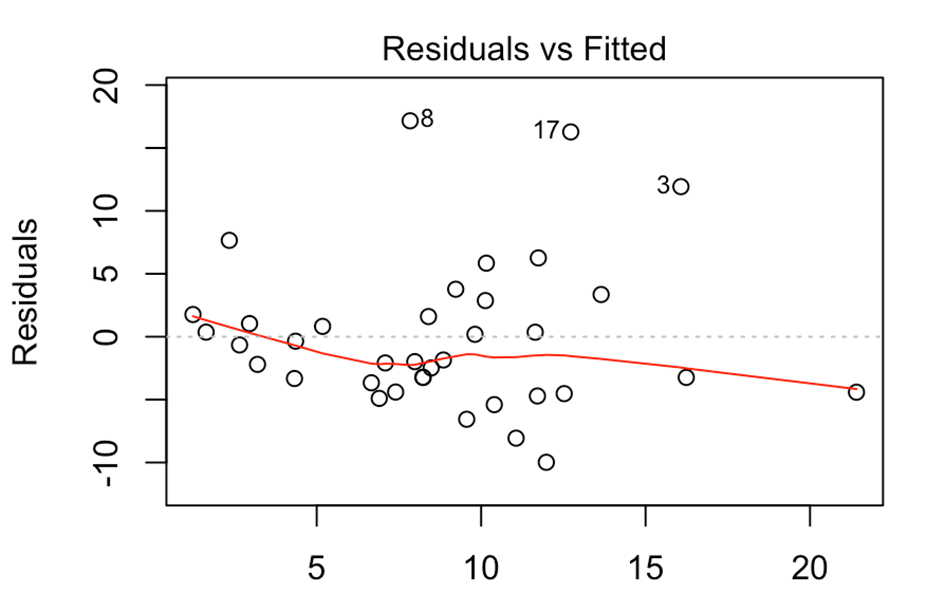}}
    \caption{Residuals \& Fitted plots for empirical model EM1, EM2, EM3}
    \label{fig:residualsPlot}
\end{figure}

We used residual plots to investigate if our metric data is suitable for linear regression analysis on EM1-EM3, see Figure~\ref{fig:residualsPlot}. We observed typical problems including non-linearity, outliers, and heteroskedasticity \cite{astivia2019}. One can see, all residual plots are in a non-random pattern, suggesting the presence of non-linearity. Shapiro-Wilks Normality Test \cite{shapiro1965} also rejected the presence of linearity in our empirical models.  In all residual plots, we can see some outliers (numbered) that are far away from the red line. The heteroskedasticity is presented, as the red line is not straight, the distribution of predicted values throughout the red line is not equal, and the residuals seem to increase with the fitted values increase in each residual plot.

In light of the above observations, Box-Cox Transformation was used to transform our data into normality and mitigate the effect of heteroskedasticity. Box-Cox Transformation can reduce anomalies in the data to ensure the usual assumption for a linear model hold \cite{box1964analysis}. For a given set of dependent and independent variables, it identifies the right specification by transforming the dependent variable \cite{velez2015new,box1964analysis}. Let $y = (y_1,y_2,...,y_n)^{'}$ be the given data on which the Box-Cox Transformation is to be applied, the Box-Cox transformation on $y$ is:

\[y \Rightarrow (y^\lambda - 1) / \lambda \]

 $y^\lambda$ is the $\lambda$-transformed data \cite{box1964analysis}. The maximum
likelihood estimate of $\lambda$ is computed to identify the transformation for the right specification \cite{box1964analysis}. All values of $\lambda$ are considered to select the optimal value for the given data $y$ \cite{velez2015new,box1964analysis}. The optimal value is the one that can result in the best approximation for the normal distribution \cite{velez2015new,box1964analysis}.

\begin{figure}[H]
    \centering
    \subfigure[ \small EM1]{\includegraphics[width=0.33\textwidth]{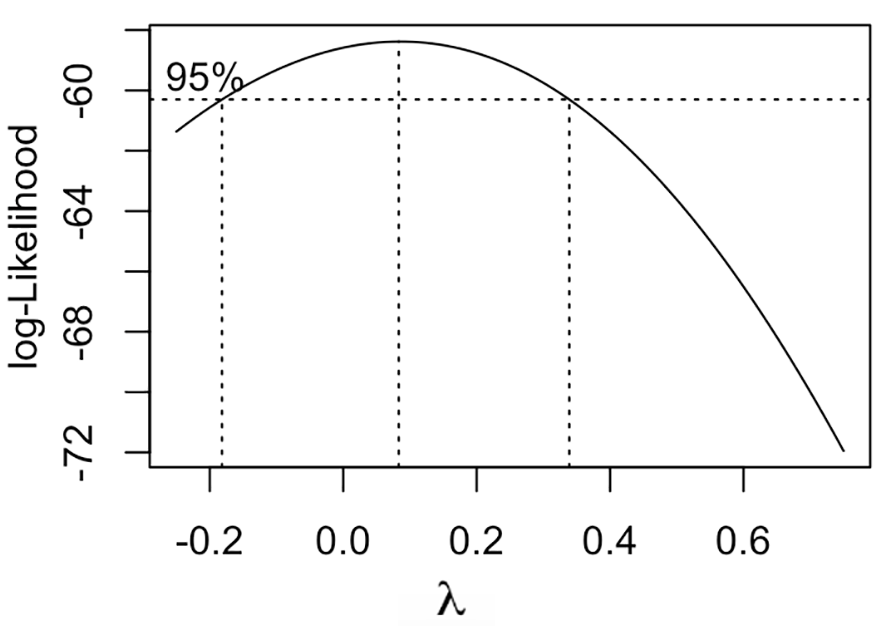}} 
    \subfigure[EM2]{\includegraphics[width=0.33\textwidth]{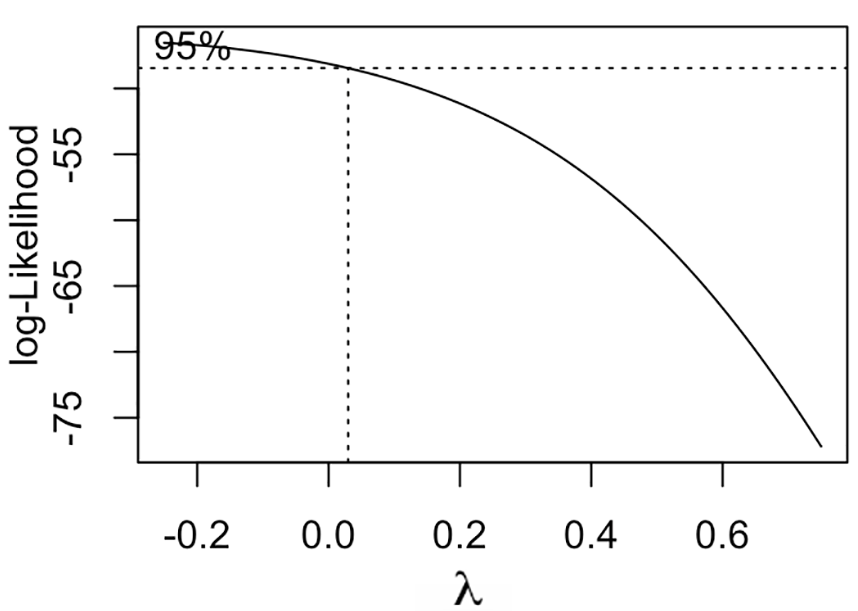}} 
    \subfigure[EM3]{\includegraphics[width=0.33\textwidth]{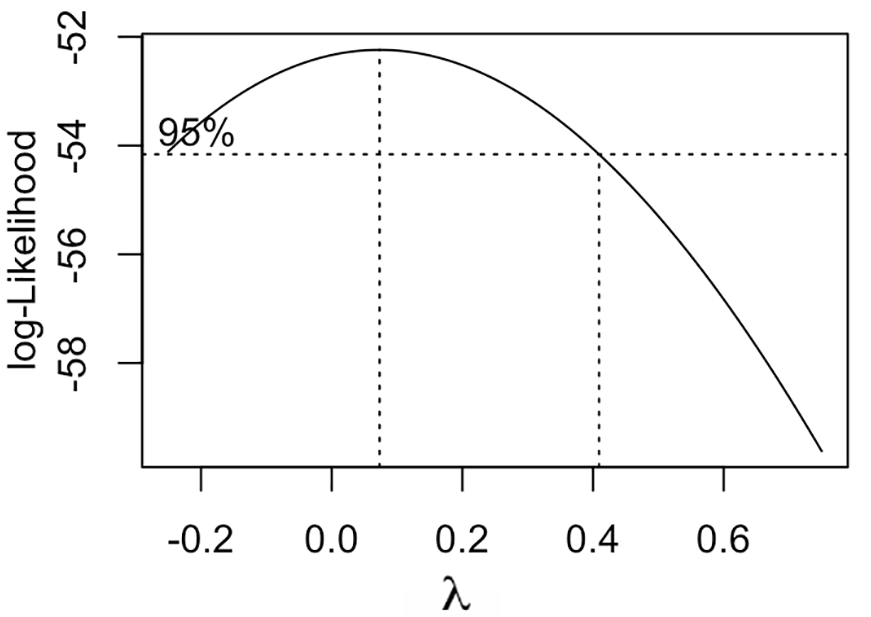}}
    \caption{Box-Cox Transformation: log-likelihood as a function of $\lambda$ values}
    \label{fig:boxcox_transform}
\end{figure}

In our paper, we used R function boxcox()\footnote{https://www.rdocumentation.org/packages/EnvStats/versions/2.4.0/topics/boxcox} to estimate $\lambda$ for our empirical model EM1-EM3. Figure \ref{fig:boxcox_transform} plots the log-Likelihood as a function of $\lambda$ values for each model. We computed that 0.083, -0.25, 0.073 are optimal $\lambda$ values that maximize the log-likelihood for EM1, EM2, and EM3. Hence, with Box-Cox Transformation, our empirical models for multiple regression analysis were defined as follows:

\begin{itemize}
    \item EM1. ((Product quality)$^{0.083} - 1 ) / 0.083$ = $\beta_{Q0} + \beta_{Q1}$(Test automation maturity) +  
    $\beta_{Q2}$(Product size) +
    $\beta_{Q3}$(Product complexity) +
    $\beta_{Q4}$(Product popularity) +
    $\beta_{Q5}$(Product age) +
    $\beta_{Q6}$(Team size) +
    $\beta_{Q7}$(Integration frequency)
    
    \item EM2. ((Test automation effort)$^{-0.25} - 1 )/ - 0.25 $ = $\beta_{C0} + \beta_{C1}$(Test automation maturity) + 
    $\beta_{C2}$(Product quality) + 
    $\beta_{C3}$(Product size) + 
    $\beta_{C4}$(Product complexity) +
    $\beta_{C5}$(Product age) +
    $\beta_{C6}$(Team size) +
    $\beta_{C7}$(Integration frequency)
    
    \item EM3. ((Release cycle)$^{0.073} - 1 ) / 0.073$ = $\beta_{R0} + \beta_{R1}$(Test automation maturity) +
    $\beta_{R2}$(Product quality) +
    $\beta_{R3}$(Test automation effort) +
    $\beta_{R4}$(Product size) +
    $\beta_{R5}$(Product complexity) +
    $\beta_{R6}$(Team size) +
    $\beta_{R7}$(Integration frequency)
\end{itemize}

\section{Results}
\label{Sec:results}

\noindent 
We present our findings in three aspects (as defined in our conceptual model): Effect of test automation maturity on product quality, Effect of test automation maturity and product quality on test automation effort, and Effect of test automation maturity and product quality on release cycle. Each aspect's findings are described below.

\subsection{Effect  of  test  automation  maturity  on  product  quality} 
\label{sec:em1_results}
\noindent Table \ref{tab:em1_results} presents the regression result of EM1. \textbf{Our hypothesis ``H1. \HoneContent" was supported.} As shown in Table \ref{tab:em1_results},  even though under the control of many other variables, ``Test automation maturity" (coefficient = -5.070e-02, p-value = 0.000624) is highly significantly associated with ``Defect density" (that measures ``Product quality"). The negative association here indicates that, 
higher levels of test automation maturity are highly significantly associated with lower defect density - higher product quality.

Moreover, we found that, the effect of test automation maturity is more significant than product size, product complexity, product popularity, product age, team size, and integration frequency on product quality. This can be seen from Table \ref{tab:em1_results},  among all variables, ``Test automation maturity'' is most significantly associated with ``Defect density''. Besides, we found that, ``Project age'' (coefficient = 2.790e-04, p-value = 0.056590) has a positive association with ``Defect density'' suggesting that old projects exhibit high defect density - lower product quality.

\begin{table}[ht]
\small
\centering
\caption{Regression results of EM1: Defect density (meas.Product quality) as dependent variable} \label{tab:TA_context}
\label{tab:em1_results}
\begin{tabular}{ l  l l l l} \hline
    & Estimate & Std. Error & statistic & p-value\\\hline
    
    (Intercept) & 7.000e-01 & 1.069e+00  &  0.655 &  0.517679 \\ 
    Test automation maturity & -5.070e-02 &  1.322e-02 &  -3.835 & 0.000624 *** \\
    Product size & -3.070e-07 & 6.437e-07 & -0.477 &  0.637045 \\
    Product complexity & 3.636e-02 & 2.708e-01 & 0.134 &  0.894115\\
    Product popularity & 1.094e-05 &  1.729e-05 &   0.633 &  0.531921 \\ 
    Project age &  2.790e-04 & 1.405e-04 &  1.986 &  0.056590 . \\
    Team size &  3.435e-02 & 6.617e-02 &  0.519 & 0.607657 \\
    Integration frequency & 1.788e-04 & 3.030e-04 &   0.590 &  0.559683
    
    \\ \hline
      \multicolumn{5}{l}{Signif. codes:  0 ‘***’ 0.001 ‘**’ 0.01 ‘*’ 0.05 ‘.’ 0.1 ‘ ’ 1}  \\\hline
\end{tabular}
\end{table}

\subsection{Effect of test automation maturity and product quality on test automation effort}
\noindent Table \ref{tab:em2_results} presents the regression result for EM2. \textbf{Our hypothesis ``H2. \HtwoContent'' was rejected}. As shown in Table \ref{tab:em2_results}, even though ``Test automation maturity'' has a positive coefficient (1.503e-03), it has no significant association (p-value =0.8611) with dependent variable ``Test automation effort''. \textbf{The regression results of EM2 also rejected our hypothesis ``H3. \HthreeContent".} As shown in Table \ref{tab:em2_results}, there was no significant association between ``Defect density'' and ``Test automation effort'', as the p-value of ``Defect density'' (0.8296) is close to 1.  The above results indicate that, increased test automation effort caused by improving the level of test automation maturity and product quality is not statistically evidenced.

Moreover, we found that, the effect of product complexity and team size is more significant than the effect of test automation maturity and product quality on test automation effort. This can be seen from Table \ref{tab:em2_results}, ``Product complexity'' (p-value = 0.0508) and ``Team size'' (p-value =  0.0449) have a significant association with dependent variable ``Test automation effort'', while ``Test automation maturity'' (p-value = 0.8611) and  ``Defect density''(p-value = 0.8296) do not have. ``Product complexity'' (coefficient =  -2.485e-01) has a significant negative association with  ``Test automation effort'',  suggesting that, more complex products cost less test automation effort. ``Team size'' (coefficient = 5.299e-02) has a significant positive association with  ``Test automation effort'', suggesting that, larger teams spend more test automation effort.

\begin{table}[ht]
\small
\centering
\caption{Regression results of EM2: Test automation effort as dependent variable}
\label{tab:em2_results}
\begin{tabular}{ l  l l l l} \hline
     & Estimate & Std. Error & statistic & p-value\\\hline
    
     (Intercept)  &  1.222e+00 & 6.188e-01 & 1.974  & 0.0579 . \\
     Test automation maturity &  1.503e-03 & 8.518e-03  &  0.176 &  0.8611 \\
     Defect density (meas.Product quality) & -2.867e-02 & 1.320e-01 & -0.217  & 0.8296  \\
     Product size & 1.810e-07 & 3.201e-07 &  0.565  & 0.5761 \\ 
     Product complexity & -2.485e-01 & 1.219e-01 & -2.038 & 0.0508 . \\
     Product age   &  4.323e-05 & 7.761e-05  & 0.557 &  0.5818  \\
     Team size & 5.299e-02 & 2.529e-02 & 2.096  & 0.0449 *\\
     Integration frequency & 8.549e-05 & 1.233e-04 & 0.693 &  0.4936 \\\hline
      \multicolumn{5}{l}{Signif. codes:  0 ‘***’ 0.001 ‘**’ 0.01 ‘*’ 0.05 ‘.’ 0.1 ‘ ’ 1}  \\\hline
\end{tabular}
\end{table}

\subsection{Effect of test automation maturity and  product quality on release cycle}

\noindent Table \ref{tab:em3_results} presents the regression result of EM3. \textbf{Our hypothesis ``H4. \HfourContent" was rejected.} One can see that, ``Test automation maturity'' (coefficient =4.135e-02, p-value = 0.01891) presents a significant positive association with dependent variable ``Number of releases''. This suggests that, achieving higher levels of test automation maturity have a positive effect on release cycle acceleration.

\begin{table}[ht]
\small
\centering
\caption{Regression results of EM3: Number of releases (meas.Release cycle) as dependent variable}
\label{tab:em3_results}
\begin{tabular}{ l  l l l l} \hline
    & Estimate & Std. Error & statistic & p-value\\\hline
    (Intercept) &  -1.194e+00 & 1.340e+00  & -0.891  &  0.38024 \\
    Test automation maturity & 4.135e-02 & 1.663e-02 &  2.486 & 0.01891 *  \\
    Defect density (meas.Product quality) &  8.913e-01 & 2.531e-01  & 3.521   & 0.00144 ** \\
    Test automation effort & 7.101e-03 & 1.031e-02  & 0.689 &  0.49643  \\
    Product size & 5.973e-08 & 7.804e-07  & 0.077 &  0.93951 \\
    Product complexity  & 9.987e-03  & 2.805e-01  &  0.036 & 0.97184  \\
    Team size & -5.701e-02 & 6.895e-02  & -0.827  &  0.41515 \\
    Integration frequency & 3.453e-04 & 3.271e-04 &  1.056 &  0.29989 \\\hline

      \multicolumn{5}{l}{Signif. codes:  0 ‘***’ 0.001 ‘**’ 0.01 ‘*’ 0.05 ‘.’ 0.1 ‘ ’ 1}  \\\hline
\end{tabular}
\end{table}

\textbf{Our hypothesis ``H5. \HfiveContent'' was supported.} As shown in Table \ref{tab:em3_results}, ``Defect density'' (coefficient= 8.913e-01, p-value=0.00144) has a significant positive association with dependent variable ``Number of releases''. This suggests that the direct effect of product quality on release time is negative - better product quality exists in longer release cycles.

\textbf{The regression result of EM3 rejected our hypothesis “H6. \HsixContent”.} We did not identify a significant association between ``Test automation effort''(p-value = 0.49643) and ``Number of releases', see Table \ref{tab:em3_results}. The result reveals that test automation effort does not seem to impact release time.

\section{Discussion}
\label{Sec:disc}
\noindent We present summary and discussion of study findings in Section \ref{sec:SumaaryStudyResults}, explore the implications for researchers in Section \ref{sec:implications_research} and implications for practitioners in Section \ref{sec:implications_pracitioner}, and examines limitations in Section~\ref{sec:limitation}.

\subsection{Summary and discussion of study findings}
\label{sec:SumaaryStudyResults}
\noindent This paper intends to answer a research question ``\OurRQ". To answer our research question, we built a conceptual model, using literature, to hypothesize the relationships between Test automation maturity, Product quality, Test automation effort, and Release cycle. We studied 37 open source java projects to test our conceptual model.  Our study findings are formulated in three aspects (as defined in our conceptual model): Effect of test automation maturity on product quality, Effect of test automation maturity and product quality on test automation effort, and Effect of test automation maturity and product quality on release cycle. We first discuss our findings in each aspect and next answer our research question.  

\paragraph{Effect of test automation maturity on product quality} We found that, higher levels of test automation maturity (assessed by standard best practices in the literature) are highly significantly associated with higher product quality in the CI context of observed open source projects. This finding suggests that standard best practices in the literature can ensure the effectiveness of test automation in detecting defects and thus assure product quality in the CI context of open source projects. Our finding is consistent with findings made by prior scholars \cite{wang2020,puri2015test,hilton2017trade}, who reported that better product quality can be achieved after adopting certain standard best practices for test automation maturity improvement in the CI context. However, our study examined a boarder range of standard best practices, as it was based on extensive literature reviews for standard best practices of test automation in our early studies  \cite{wang2019,icsoft20}.

To extend prior research on this topic, when analyzing the relationship between test automation maturity and product quality, we involved several control variables: product size, product complexity, product popularity, product age, team size, and integration frequency. Our study is the first attempt to evidence that, the effect of test automation maturity is more significant than product size, product complexity, product popularity, product age, team size, and integration frequency on product quality in the CI context of open source projects. It again confirms the importance of improving the level of test automation maturity in the CI context of open source projects and also the positive impact of standard best practices on that. Furthermore, we found that, old projects exhibit lower product quality. This finding conflicts with the observation of prior scholars \cite{harter2000}, who examined that, older products have a lower chance of exposure  to defects. The difference might come from the approach to compute defect density (we used post-release defects while they used pre-release defects), types of projects (we observed open sources projects while they observed closed source projects), or varied project technologies (we selected projects that are written in Java and use CI tools while their sampled projects were different). Besides, many studies~\cite{cotroneo2016bugs,ibrahim2012relationship,au2009virtual} have observed that, larger teams may insert more defects to a software product because more people means a higher occurrence chance of defects. Yet, our study did not find a significant association between team size and product quality. The possible explanation is that, compared to prior studies, our study included more variables - as such, the effect of team size on product quality is not that significant under the control of other variables.

\paragraph{Effect of test automation maturity and product quality on test automation effort}
Our study did not find a significant association between test automation maturity/product quality and test automation effort in the CI context of observed open source projects, when controlling for product size, product complexity, product age, team size, and integration frequency. This reveals that, increased test automation effort caused by improving the level of test automation maturity and product quality is not evidenced. Our finding conflicts with prior studies' viewpoint \cite{lin2020test,hilton2017trade,lee2012software,williams2009,kasurinen2010software}, which states that, at higher levels of test automation maturity and product quality, there is increased effort to develop and execute rigorous automated tests. Our finding confirms the other viewpoint, which states the effort savings may offset the invested effort when using standard best practices to perform rigorous test automation  \cite{berner2005,kumar2016impacts,puri2015test,tosun2009}. Based on many sources \cite{graham2012experiences,garousi2010replicated,garousi2017taNot}, test automation maturity improvement is an investment that needs efforts to implement, while the return is not linear - cost savings would be achieved after a certain period of time and then increase with the time grows. That could be the explanation for the existence of prior two different viewpoints.  Studies~\cite{puri2015test,tosun2009,berner2005} observed that, effort savings (from the reuse of quality test artifacts, effective strategic planning and test prioritization, structured test automation process, the usage of development guidelines, improved SUT testability, and increased competency of test professionals) exceed the invested effort 
when test automation maturity improvement programs were fully implemented. On the contrary, studies \cite{lee2012software,williams2009} observed increased test automation effort  when a standard test process was initialized or not fully implemented, while studies~\cite{lin2020test,hilton2017trade} did not consult cost-savings with respect to the progress of test automation maturity improvement.  

Concerning control variables, we found that the effect of product complexity and team size is more significant than test automation maturity and product quality on test automation effort. This finding is not presented by prior scholars. Many prior studies \cite{staahl2017continuity,banker1992software,zhao2017impact,shahin2017continuous} observed that larger teams would spend more test automation effort. Our study evidenced that their observation is valid in the CI context of open source projects. Besides, our study identified that more complex products spend less test automation effort. This finding conflicts with the observation of prior scholars \cite{harter2000,agrawal2007}, who have empirically proofed that it would take more effort to develop and execute automated tests on more complex software products. One possible explanation is that for more complex products developers automate fewer tests due to complexity \cite{garousi2016and}. 

\paragraph{Effect of test automation maturity and product quality on release  cycle} 
Our study empirically evidenced that, higher levels of test automation maturity (assessed by standard best practices in literature) are significantly associated with shorter release cycles, controlling for product size, product complexity, team size, and integration frequency. Our finding is consistent with prior studies' observation, which views short release cycles can be enabled by improving the level of test automation maturity  \cite{wang2020,collins2012,berner2005,kumar2016impacts,puri2015test}. On the contrary, our finding conflicts with the other viewpoint, which states, release time must increase against improvements of the level of test automation maturity~\cite{kasurinen2010software,williams2009}. Such a prior viewpoint explained that, developing and executing rigorous automated tests (using standard best practices) to ensure product quality requires considerable effort, so that the elapsed time is added for product development and thus product releases are delayed  \cite{kasurinen2010software,williams2009}. Yet, in our study, the association between test automation effort and release cycle was not statistically evidenced. This suggests test automation effort does not seem to affect release time in the CI context of open source projects.

Prior studies have empirically observed that, with short release cycles, even though the testing period is short, defects were fixed early so that product quality is improved in the CI or CI like context \cite{wang2020,berner2005,puri2015test,collins2012}. Our finding conflicts with the observation of these prior studies. We observed that the direct effect of product quality on release time is negative - better product quality exists in longer release cycles - in the CI context of observed open source projects. One possible explanation is, the number of releases depends on the evolving set of features that may insert defects. Scholars \cite{greer2004software,ruhe2005art} observed that, as a software release is a collection of new and/or changed features, more new and/or changed features would enable more frequent releases; However, since features are developed in codes and ``defects are coding caused'', more new and/or changed features would insert more defects. Another potential reason could be there is a threshold for the length of release time. That is, there needs sufficient time to make releases in the CI context. Based on a case study \cite{hamdan2015quality} that measured product quality before and after applying CI in the development of a software system, it usually took some time to ``plan a release'' (define release goal, preview the product backlog, and testing a release) before publishing it to users, and hence, too short release cycles in which releases are not planned adequately may make products to be easily exposed to post-release defects (that was used to measure product quality in our research). Also, if the release time is too short, the testing period is too short so that test automation can not reach the proper deep and cover the right scope, and thus, product quality suffers \cite{shahin2017continuous,hamdan2015quality}.

\paragraph{Answers on our research question}  
Our observations suggest that, in the CI context of open source projects, a potential benefit of improving the level of test automation maturity against standard best practices in the literature is product quality improvement and release cycle acceleration, while increased test automation effort caused by improving the level of test automation maturity and product quality is not evidenced. Thus, we draw a conclusive answer for our research question: higher levels of test automation practices can lead to better product quality without increased release time and test automation effort in the CI context of open source projects.

\subsection{Implications for researchers}
\label{sec:implications_research}
\noindent From a theoretical standpoint, our results suggest several directions for theory development in this research domain. First, our study found that using standard best practices in the literature to conduct test automation can enable CI success in open source projects. Researchers in this domain could explore how to adopt each standard best practice (mentioned in our survey) more deeply. Future research can build on these results to develop test automation maturity models for the CI context. Second, we noted some different observations between prior studies and our study around some control variables. 
In particular, many studies observed that larger teams tend to insert more defects to a software product, while our study did not find a significant association between team size and product quality; Prior studies observed that more complex products need more effort to develop and execute automated tests, whereas our observation is different; Prior studies observed that older products expose fewer defects due to the increased experience of the team, while we observed that old projects exhibit lower product quality. Due to the scope of our study, we only noted different observations on those control variables and explored possible reasons leading to that. Future efforts to provide theoretical explanations for such different observations would be meaningful. Third, our study demonstrates a novel investigation of CI success from a test automation maturity based view. Prior studies have actively researched on CI process as the determinant of CI success, for example, the integration flow, building frequency, integration serialization and batching, building status communication \cite{staahl2014,ghaleb2019empirical,staahl2017continuity}. Future research could also include the view of test automation maturity as the determinant in CI success literature. 

Our study demonstrates a triangulated method, that surveyed practitioners and mined project data from multiple repositories (GitHub repository, Travis CI repository, issue track system), to study test automation practices in the CI context of open source projects. Other researchers can consult this triangulated method to study other test automation topics in the CI context, e.g., the effectiveness of test automation, execution efficiency of test automation, or productivity outcomes of adopting CI with test automation.  

In our study, we studied many variables and collected metric data on these variables: test automation maturity, product quality, test automation effort (test automation development effort and execution effort), release cycle, product size, product complexity, product age, product popularity, team size, and integration frequency. For researchers who observe similar variables in their research, our study would provide the reference on metric selection and data collection on their observed variables. We clearly documented our data collection process. This would give others a hint about how to mine relevant metric data on these variables from open source project repositories. 

Moreover, our dataset contains test automation maturity survey responses, project data, test suite, and test logs of 37 open source java projects. We have made our dataset publicly available. Our dataset could be used by researchers to study other test automation topics. For example, test suite and test logs of these projects can be used for test prioritization or test case selection related studies; Test code quality related studies can make observations on test suites of these projects; Test automation survey responses and test logs can be studied together to explore the state of art and practice of test automaton of open source projects; Quality assurance outcomes related studies can use the count of reported bugs of these projects.

\subsection{Implications for practitioners} 
\label{sec:implications_pracitioner}
\noindent 
Our study has implications for practitioners who are working on CI with test automation. Our study evidenced that, a potential benefit of improving the level of test automation maturity (using standard best practices in the literature) is product quality improvement and release cycle acceleration in the CI context of open source projects. Knowing this finding would help practitioners to assess and improve their test automation maturity towards CI success in the similar CI context within their organizations. They could assess and improve test automation practices against standard best practices mentioned in our survey. They also can collect similar metric data (used in our study) to observe how test automation maturity affect product quality, test automation effort, and release cycle in their CI context, or analyze costs and benefits of test automation/CI, i.e., good product quality and short release cycles as benefits and test automation effort as costs.

\subsection{Limitations}
\label{sec:limitation}
\noindent Like any other studies, this study also has some limitations. First, a limitation is that our dataset might not be representative of all open source projects that are adopting test automation in the CI context. We studied 37 open source java projects using Travis CI and Maven. We must select projects in the same programming language, using the same CI server tool, and using the same CI building framework, because different ones may have different performance and structure (that may bias our research results). Although we aimed to select and study the large sample of open source java projects, given our limited resource constraints, it finally did not realize in the end. In the project selection phase, we began with JTec and 20-MAD datasets that contain over 30k open source java projects in total, but in the end only 149 projects met our needs. A large number of projects were excluded as we did not have the access to those projects' private Travis CI repository and issue tracker system. In the survey distribution phase, we also lost the projects in which selected contributors did not answer our survey. Thus, we encourage future research to extend our findings by adding more datasets. For example, extending datasets with open source projects written in other popular programming languages such as Python, C++ and JavaScript, or use the other CI server tool (like Jenkins or CircleCI) and building tool (like Gradle).

Second, since our study only examines open source projects, the limitation lies in whether our study results can be generalized to other types of software development contexts that adopt test automation using similar CI tools, for example, closed source software development contexts and large-scale industrial software development contexts. By considering differences, similar experiment studies can be carried out to validate our conceptual model (Section \ref{sec:conceptualModel}) in other types of software development contexts. For example, as full-time developers are hired to work on closed source projects, estimating or calculating test automation effort using the metric Person-days is more accurate than the metric LOC used in our study.  Another example is, for complex large-scale industrial projects consisting of millions of LOCs, the control variable ``Product complexity'' may need to be measured by the integration of multiple metrics such as  Cyclomatic complexity number (used in our study), Structural coupling, and Logical coupling  \cite{sarkar2008metrics}.

Third, we used ``Increased LOC of test automation (without counting of blanks and comments)'' as the metric to measure test automation development effort, because this metric is more widely used than other metrics for open source projects (see Section \ref{sec:project_data_collection}) and  calculating the actual test automation development effort for each project is not feasible. However, we believe that future research is needed to validate our findings with the actual test automation development effort in open source projects and others (e.g., closed source projects and large-scale industrial projects).

Finally, in our conceptual model, we included control variables that affect the relationships between our observed  variables. All control variables were identified by prior scholars. Additional control variables may exist in the real industrial context but have not been identified by prior scholars. When new control variables are identified, the replication of this study is required to evaluate findings by including new control variables.

\section{Conclusion}
\label{Sec:conc}
\noindent In this paper, we have empirically studied the effect of test automation maturity (assessed by standard best practice in the literature) on product quality, test automation effort, and release cycle in the CI context of 37 open source java projects. Our study showed that, in such the CI context, a potential benefit of improving the level of test automation maturity is product quality improvement and release cycle acceleration, while increased test automation effort caused by improving the level of test automation maturity and product quality is not evidenced. Our results suggest that test automation related standard best practices defined in the literature would fit the CI context of open source projects and enable CI success. Our study evidenced some observation of prior studies, resolved the different viewpoints of prior studies, and identified novel research topics to extend the impact and scope of CI and test automation literature, see details in Section~\ref{sec:SumaaryStudyResults}~and Section~\ref{sec:implications_research}. Our recommendation to practitioners (in the similar CI context) is to utilize standard test automation best practices to improve test automation maturity   towards test automation success as well as CI success, see details in Section \ref{sec:implications_pracitioner}.

\section*{CRediT author statement}
\noindent \textbf{Yuqing Wang:} Conceptualization, Methodology, Investigation, Formal analysis, Data Curation, Writing-Original draft preparation. \textbf{Mika V. M{\"a}ntyl{\"a}:} Conceptualization, Methodology, Investigation, Writing-Review \& Editing, Supervision, Funding acquisition. \textbf{Zihao Liu:} Methodology, Investigation, Data Curation, Formal analysis, Validation. \textbf{Jouni Markkula:} Methodology, Writing-Review \& Editing, Supervision. 

\section*{Acknowledgments}
\noindent This work is supported by TESTOMAT Project (ITEA3 ID number 16032) funded by Business Finland under Grant Decision ID 3192/31/2017, and the foundation of Tauno Tönning (project ID 20210086).







\bibliographystyle{elsarticle-num-names}
\bibliography{main.bib}







\end{document}